\documentclass[reprint,10pt,aps, superscriptaddress]{revtex4-2}

\usepackage{graphicx} % Enhanced LaTeX Graphics

\usepackage[justification=justified,format=plain]{caption}

\usepackage{subfigure} % subcaptions for subfigures
\usepackage{subfigmat} % matrices of similar subfigures

\usepackage{dcolumn}           % decimal-aligned tabular math columns
\newcolumntype{.}   {D{.}{.}{-1}} % column alignedd on the point separator '.'
\newcolumntype{d}[1]{D{.}{.}{#1}} % column centered on the point separator '.'
\newcolumntype{e}   {D{E}{E}{-1}} % column centered on the exponent 'E'
\newcolumntype{E}[1]{D{E}{E}{#1}} % column centered on the exponent 'E'

\usepackage{amsmath}
\usepackage{amsfonts}
\usepackage{amssymb}
\usepackage{amsthm}

\usepackage{comment}

\usepackage{float}

\usepackage{hyperref}
\hypersetup{ colorlinks=true,
             citecolor=cyan,
             linkcolor=darkgray,
             urlcolor=teal,
             breaklinks=true,
             bookmarksnumbered=true,
             bookmarksopen=true,
}

\usepackage[english]{cleveref}
\usepackage{physics}
\usepackage[T1]{fontenc}
\usepackage{helvet}
\usepackage{subcaption}
\usepackage{array}
\usepackage{tikz}
\usetikzlibrary{shapes.geometric, arrows, positioning, arrows.meta,quantikz2}
\usepackage{multirow}
\usepackage{graphicx}

\begin{document}

\title{Energetics of Trapped-Ion Quantum Computation}
\date{\small{\today}}

\author{Francisca Góis}
\affiliation{PQI -- Portuguese Quantum Institute, Portugal}

\author{Marco Pezzutto}
\affiliation{PQI -- Portuguese Quantum Institute, Portugal}
\affiliation{Physics of Information and Quantum Technologies Group, Centro de Física e Engenharia de Materiais Avançados (CeFEMA), Portugal}
\affiliation{LaPMET -- Laboratory of Physics for Materials and Emerging Technologies, Portugal}

\author{Yasser Omar}
\affiliation{Instituto Superior Técnico, Universidade de Lisboa, Portugal}
\affiliation{PQI -- Portuguese Quantum Institute, Portugal}
\affiliation{Physics of Information and Quantum Technologies Group, Centro de Física e Engenharia de Materiais Avançados (CeFEMA), Portugal}
\affiliation{LaPMET -- Laboratory of Physics for Materials and Emerging Technologies, Portugal}
\affiliation{Quantum Green Computing, Ltd., Portugal}

%%%%%%%%%%%%%%%%%%%%%%%%%%%%%%%%%%%%%%%%%%%%%%%%%%%%%%%%%%%%%%%%%%%%%%
% ABSTRACT & KEYWORDS
%%%%%%%%%%%%%%%%%%%%%%%%%%%%%%%%%%%%%%%%%%%%%%%%%%%%%%%%%%%%%%%%%%%%%%
%\vspace{-2cm}
%%%%%%%%%%%%%%%%%%%%%%%%%%%%%%%%%%%%%%%%%%%%%%%%%%%%%%%%%%%%%%%%%%%%%%
%     File: Paper_abstr.tex                               %
%     Tex Master: Paper.tex                               %
%                                                                    
%%%%%%%%%%%%%%%%%%%%%%%%%%%%%%%%%%%%%%%%%%%%%%%%%%%%%%%%%%%%%%%%%%%%%

%%
%% Abstract
%%
\begin{abstract}

    \noindent 
    The question of the energetic efficiency of quantum computers has gained increasing attention recently. A precise understanding of the resources required to operate a quantum computer with a targeted computational performance and how the energy requirements can impact the scalability is still missing. 
    In this work, one implementation of the quantum Fourier transform algorithm in a trapped-ion setup was studied. The main focus was to obtain a theoretical characterization of the energetic costs of quantum computation, based on actual experimental measurements performed on a similar trapped-ion setup.
    The energetic cost of the computation was estimated by analyzing the components of the setup and all the steps involved, from the cooling and preparation of the ions to the execution of the algorithm and readout of the result. In the Noisy Intermediate-Scale Quantum regime, a potential scaling of the energetic costs was argued and used to find a possible threshold for an energetic quantum advantage against state-of-the-art classical supercomputers. Remarkably, this threshold appears to be lower than the one for which computational time advantage is expected.
\\
%%
%% Keywords (max 5)
%%

\end{abstract}

\maketitle
%%%%%%%%%%%%%%%%%%%%%%%%%%%%%%%%%%%%%%%%%%%%%%%%%%%%%%%%%%%%%%%%%%%%%%
% INTRODUCTION
%%%%%%%%%%%%%%%%%%%%%%%%%%%%%%%%%%%%%%%%%%%%%%%%%%%%%%%%%%%%%%%%%%%%%%
%%%%%%%%%%%%%%%%%%%%%%%%%%%%%%%%%%%%%%%%%%%%%%%%%%%%%%%%%%%%%%%%%%%%%%
%     File: Paper_intro.tex                               %
%     Tex Master: Paper.tex                               %
%                                                                    
%%%%%%%%%%%%%%%%%%%%%%%%%%%%%%%%%%%%%%%%%%%%%%%%%%%%%%%%%%%%%%%%%%%%%

\section{Introduction}
\label{sec:intro}

Quantum computation aims at harnessing the laws of quantum mechanics to obtain a quantum advantage in solving computational problems that are not tractable by classical computers in a feasible time. 
The development of quantum computers is being supported by a growing investment of resources. 
Although the scalability of these technologies remains a challenge, their development is driven by the potential applications of the power of quantum computation, from optimization and cryptoanalysis, to finance and computational chemistry. 
The experimental efforts have been focused on achieving increasingly better time performances and larger numbers of qubits. 
The question of the energy performance of quantum computers, however, has been playing a secondary role: a precise understanding of the connection between a targeted computational performance and the required resources, in particular the energetic consumption, is still missing. 
%An understanding of the energetic consumption of quantum computers is important to bring the awareness of energetic concerns~\cite{inalexia} in the research on the various qubit technologies and their scalability. 
Is important to bring the awareness of energetic concerns in the research on the various qubit technologies and their scalability.
This is even more relevant in view of a possible quantum advantage from the energy perspective, which provides a new argument in support of the development of quantum computers.

The topic of energetics of quantum technologies has received increasing attention recently: there have been investigations on the energetic costs of different quantum control protocols \cite{en-control}, quantum measurements \cite{en-meas}, quantum logic gates \cite{en-min, en-eneff, en-sqgate}, quantum information technologies applied to classical computation \cite{sagar, advantage, dylan}, and efforts to estimate the energy costs of different elements of quantum computers \cite{en-datacenters, en-error}. With the goal of determining a threshold for a ``green'' quantum advantage, estimates for the energy costs of different quantum hardware platforms were compared with classical implementations \cite{green}. Many works addressed specifically the minimization of the energetic footprint of quantum computers, looking for key elements where the resource costs can be optimized \cite{green,en-datacenters,main,asiani}. Furthermore, in~\cite{Stevens_2025} the authors investigate the energetic cost of individual quantum gates through a microscopic Hamiltonian approach, obtaining a lower bound of the energy per gate for a given accepted error threshold, and argue that the energetic requirements may be reduced at the expense of increasing computational complexity. Furthermore, \cite{Ariane2026} presents a discussion of the advantages and inconveniences of different current quantum computing platforms from an energetic standpoint.

A truly comprehensive study of the energetics of quantum computation is thus paramount, addressing all its components, namely the energetic costs of the execution of the quantum gates/algorithms, of the quantum data buses, of the baseline costs of running the experimental setup (e.g., fields and lasers generating traps, vacuum, cryogenics, etc.), and of the classical control of the experiment. This research agenda should include the costs of generating non-trivial initial states, interconnecting different quantum processors, etc., and establish benchmarks to assess the energetic performance of quantum machines. Furthermore, the energetic costs will naturally depend on the chosen platform, requiring dedicated studies.

In this work, we study an implementation of the Quantum Fourier Transform (QFT) in a trapped-ion system \cite{versatile}. Our energetic analysis is built upon data acquired through direct measurements in a similar trapped-ion setup,  presented in~\cite{sagar}, and performed in the first place to explore the usage of quantum technologies for classical information processing. Similar studies investigate the energetic performance of the QFT implemented in quantum processors based on Rydberg atoms~\cite{Oscar2026}, spin qubits in semiconductors~\cite{Joao2026} and superconducting CAT qubits~\cite{Pedro2026}.

The implementation of the QFT algorithm was simulated and validated through comparison with the theoretical ideal previsions, using a selected metrics of performance. The quantum noise and its impact on the performance of the algorithm were studied including three different noise models in the simulation. A theoretical characterization of the energetic consumption of the computation was obtained by analyzing the different components of the experimental setup and the various stages involved. Additionally, a potential scaling of the energetic costs was argued and used to find a possible threshold for an energetic quantum advantage against state-of-the-art classical supercomputers.

%%%%%%%%%%%%%%%%%%%%%%%%%%%%%%%%%%%%%%%%%%%%%%%%%%%%%%%%%%%%%%%%%%%%%%
% BACKGROUND
%%%%%%%%%%%%%%%%%%%%%%%%%%%%%%%%%%%%%%%%%%%%%%%%%%%%%%%%%%%%%%%%%%%%%%
%%%%%%%%%%%%%%%%%%%%%%%%%%%%%%%%%%%%%%%%%%%%%%%%%%%%%%%%%%%%%%%%%%%%%%
%     File: Paper_QFT.tex                               %
%     Tex Master: Paper.tex                               %
%                                                                    
%%%%%%%%%%%%%%%%%%%%%%%%%%%%%%%%%%%%%%%%%%%%%%%%%%%%%%%%%%%%%%%%%%%%%

\section{Quantum Fourier Transform with Trapped Ions}
\label{sec:backg}

In the experiment that was studied \cite{versatile}, a coherent QFT was realized on three qubits using an approach based on microwave-driven trapped ions.

%%%%%%%%%%%%%%%%%%%%%%%%%%%%%%%%%%%%%%%%%%%%%%%%%%%%%%%%%%%%%%%%%%%%%%
\subsection{Microwave-driven Trapped Ions}

Trapped ions are among the most promising platforms to build quantum hardware for quantum computing. 
In the experiment, three $^{171}$Yb$^+$ ions were confined in a linear Paul trap. The qubit states are encoded in the two hyperfine levels of the ground state $S_{1/2}$. The two states are connected by a magnetic dipole resonance near $\omega_{\text{Res}} = 2 \pi \times 12.6$ GHz, which is controlled with microwave radiation.

In quantum information processing with trapped ions, the qubit transitions are generally driven by intricate laser systems. 
Simpler methods can be created using long-wavelength radiation, such as microwaves, employing an additional magnetic field gradient applied along the axis of the ion trap. The magnetic field gradient shifts the qubit resonances, making the ions distinguishable in frequency space, and introduces a coupling between the internal and motional states, a magnetic gradient induced coupling (MAGIC) \cite{magic}. 
Due to this coupling, through Coulomb interaction, when one ion changes its equilibrium position, the position and internal energy of the neighbouring ions are also changed and an indirect spin-spin coupling is established \cite{khromova}.

The spin-spin interaction of a chain of $n$ ions in a magnetic field gradient is described by the Hamiltonian

\begin{equation}
    \mathcal{H}_{\text{I}} = -\frac{\hbar}{2} \sum^n_{k,l=1, k\neq l} J_{kl} \\\ \sigma_z^{(k)} \sigma_z^{(l)},
    \label{eq:hamiltonian}
\end{equation}
where $\sigma_z^{(k)}$ is the $z$ Pauli matrix on the subspace of ion $k$ and $J_{kl}$ are the spin-spin coupling constants between two different ions $k$ and $l$. 

In the experimental setup, the magnetic field gradient is $b = 19$ T/m. The resulting addressing frequencies of the ions are $\omega_1 = 2\pi \times 12.645$ GHz, $\omega_2 = 2\pi \times 12.648$ GHz and $\omega_3 = 2\pi \times 12.651$ GHz, which corresponds to a frequency separation of about $3.1$ MHz \cite{khromova}. The mutual coupling constants were measured experimentally, giving the results $J_{12} = 2\pi \times 34$ Hz, $J_{23} = 2\pi \times 39$ Hz and $J_{13} = 2\pi \times 27$ Hz \cite{versatile}.

The qubits can be prepared in any state through the interaction with the microwave field for a certain time $t$, implementing a rotation of angle

\begin{equation}
    \theta = \Omega t,
    \label{eq:theta}
\end{equation}
where $\Omega \approx 2\pi \times 50$ kHz is the Rabi frequency. %, which measures the coupling strength of the internal state of an ion to the electromagnetic radiation. 
Each single-qubit operation is thus mapped to the respective microwave pulse, characterized by a specific frequency, tuned to the resonance frequency of the desired qubit, and by a specific time duration, associated with the rotation angle of the operation.

Before an algorithm can be performed in the quantum computer, the ions have to be cooled, which is achieved by reducing their kinetic energy to bring the ions close to their motional ground state. The cooling of the ions is provided by Doppler cooling and sideband cooling. 
Then, the ions are prepared in the ground state, followed by the initialization with the desired input state for the algorithm.
Finally, after the sequence of computational gates is applied with the microwave pulses, the output state can be measured \cite{sagar,khromova}. 
These operations are carried out just one time for a sequence of computational gates and are implemented with a combination of microwave pulses and laser lights near $369.5$ nm and $935.2$ nm.

%%%%%%%%%%%%%%%%%%%%%%%%%%%%%%%%%%%%%%%%%%%%%%%%%%%%%%%%%%%%%%%%%%%%%%
\subsection{Quantum Fourier Transform}

The QFT performs a Fourier transform of the amplitudes of a quantum state and it can be applied to determine their periodicity. It is defined as a linear unitary operator acting on an orthonormal basis $\{\ket{0},...,\ket{N-1}\}$ as

\begin{equation}
    \ket{j} \mapsto \frac{1}{\sqrt{N}} \sum_{k=0}^{N-1} e^{2 \pi i j k /N} \ket{k}.
\label{eq:qft}
\end{equation}
$N=2^n$, with $n$ the number of qubits.

The action of the QFT on an arbitrary state is given by:

\begin{equation}
    \begin{split}
    \sum_{j=0}^{N-1} x_j \ket{j} &\mapsto \sum_{k=0}^{N-1} \bigg(\frac{1}{\sqrt{N}} \sum_{j=0}^{N-1} x_j  e^{2 \pi i j k /N} \bigg) \ket{k} \\
    &= \sum_{k=0}^{N-1} X_k \ket{k}.
    \end{split}
\label{eq:qfteqsup}
\end{equation}

The QFT implementation that was studied is adapted for systems described by Ising-type Hamiltonians of the form \eqref{eq:hamiltonian}. It is implemented by resorting to the following gate set:

\begin{itemize}
    \item Rotation, $R_k(\theta)$: Implemented by a microwave pulse resonant with the spin-flip transition frequency of each ion.
    \begin{equation}
        R_k(\theta) = e^{-i \frac{\theta}{2} \sigma_x^{(k)}}
    \end{equation}
        
    \item Phase gate, $\Phi_k(\phi)$:
    \begin{equation}
        \Phi_k(\phi) = e^{-i \phi \sigma_z^{(k)}}
    \end{equation}
    
    \item Phased rotation, $R_k(\theta, \phi)$: Obtained with a phase shift of the driving microwave field implementing the rotation, with $R_k(\theta, 0) = R_k(\theta)$.
    \begin{equation}
        \begin{split}
        R_k(\theta, \phi) & \equiv \Phi(\phi/2) R(\theta) \Phi(-\phi/2) \\ &= e^{-i \frac{\theta}{2} (\sigma_x \cos \phi + \sigma_y \sin \phi)}
        \end{split}
        \label{eq:phasedrotation}
    \end{equation}       

\end{itemize}

\begin{figure*}[ht!]
    \[
    \begin{array}{c}
        \centering
            \begin{quantikz}
                & \gate{H} &  \gate[3]{U(T_1)} &  & \gate[3]{U\Big(\frac{T_3}{2}\Big)} & \gate{R\Big(\pi, \frac{\pi}{2}\Big)} & \gate[3]{U\Big(\frac{T_3}{2}\Big)} & \gate{R\Big(\pi, \frac{27\pi}{16}\Big)} & \swap{2} &  \\
                &  & \ghost{U(T_1)} & \gate{R(\pi,0) R\Big(A_1, \frac{3\pi}{4}\Big)} & \ghost{U\Big(\frac{T_3}{2}\Big)} &  & \ghost{U\Big(\frac{T_3}{2}\Big)} & \gate{R\Big(A_2, \frac{3\pi}{4}\Big)} & &  \\
                &  & \ghost{U(T_1)} & \gate{R\Big(\pi, \frac{13\pi}{16}\Big)} & \ghost{U\Big(\frac{T_3}{2}\Big)} &  & \ghost{U\Big(\frac{T_3}{2}\Big)} & \gate{R\Big(\frac{\pi}{2}, \frac{3\pi}{2}\Big)} &  \targX{} &  \\
            \end{quantikz}
    \end{array}
    \]
    \caption{Circuit that realizes the quantum Fourier transform based on single-qubit rotations $R(\theta,\phi)$ and free evolutions $U(T)$, performing the optimized sequence of operations \eqref{eq:Uopt}.}
    \label{fig:optcircuit}
\end{figure*}
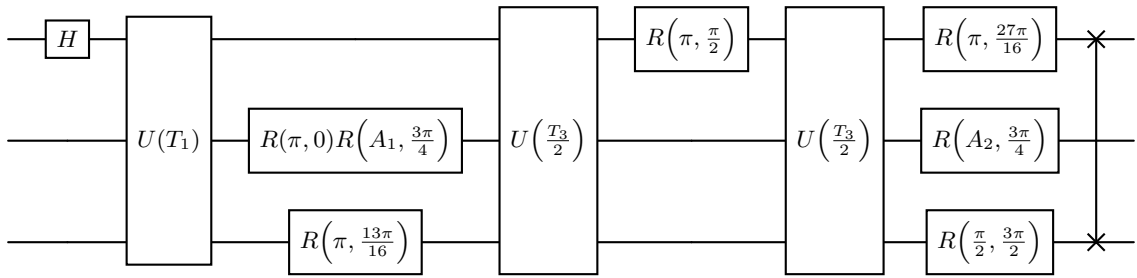

\begin{itemize}
    \item Entangling gate, $U_{kl}(t)$: Obtained from the free evolution of the ions $k$ and $l$ for a certain evolution time $t$, with all the other ions being uncoupled.
        \begin{equation}
            U_{kl}(t) = e^{iJ_{kl}t/2 \sigma_z^{(k)}\sigma_z^{(l)}}
        \end{equation}
        $J_{kl}$ represents the spin-spin coupling strength for qubits $k$ and $l$. When all the qubits are fully coupled, the entangling gate becomes the natural time evolution of the system with the Hamiltonian \eqref{eq:hamiltonian},
    
        \begin{equation}
            U(t) = e^{-i \mathcal{H}_{\text{I}} t/ \hbar}  
            \label{eq:uentangling}           
        \end{equation} 
    
    \item Hadamard gate, $H_k$: Decomposed into two rotations, obtained (up to a phase) with two pulses with a phase difference of $-\pi/2$.
    \begin{equation}
        H_k = i R(\pi/2,-\pi/2) \ R(\pi,0)
        \label{eq:hadamardecomp}
    \end{equation}

\end{itemize}

The indices $k$ indicate to which qubit the operations are applied.

The gate sequence that implements the QFT up to a global phase, derived in \cite{simplified} and optimized in \cite{versatile}, is

\begin{equation}
    \begin{split}
        U_{\text{QFT}}^{\text{opt}} = & R_2\Big(A_2, \frac{3\pi}{4}\Big) R_3\Big(\frac{\pi}{2}, \frac{3\pi}{2}\Big) R_1\Big(\pi, \frac{27\pi}{16}\Big) \\
        & U\Big(\frac{T_3}{2}\Big) R_1\Big(\pi, \frac{\pi}{2}\Big) U\Big(\frac{T_3}{2}\Big) R_2\Big(A_1, \frac{3\pi}{4}\Big)\\
        & R_2(\pi,0) R_3\Big(\pi, \frac{13\pi}{16}\Big) U(T_1) H_1.
    \end{split}
    \label{eq:Uopt}
\end{equation}
The experimental parameters of the setup give the evolution times $T_1 = 3.69$ ms, $T_2 = 0.22$ ms, and $T_3 = 4.87$ ms. The pulse areas take the values $A_1 = 0.686\pi$ and $A_2 = 0.716\pi$.

At the end of the sequence \eqref{eq:Uopt}, an additional swap gate interchanges the state of qubits 1 and 3, $SWAP_{13}$, which is implemented experimentally by just relabeling the qubits.
 
The complete circuit that performs the QFT is presented in figure \ref{fig:optcircuit}.

Since the Rabi frequency is about $2\pi \times 50$ kHz, the duration of the single qubit rotations can be neglected. The total duration of the QFT gate sequence arises from the conditional evolution times, $t_{\text{QFT}} = T_1 + \frac{T_3}{2} + \frac{T_3}{2} = 8.56$ ms.

The experimental setup is affected by noise, mainly due to fluctuations of the electromagnetic fields of the trap, limiting the coherence time of the qubits to $200$ $\mu$s \cite{khromova}. 
Dynamical decoupling (DD) techniques can be used to extend the coherence time by applying periodic sequences of instantaneous control pulses \cite{phdthesis}.
A total of $60$ DD $\pi$ pulses were applied to each qubit during the implementation of the QFT. Since the DD pulses are implemented by single qubit rotations of $\pi$, their duration can also be neglected.

To validate the gate sequence \eqref{eq:Uopt} in performing a QFT, a simulation of the circuit was created and some noise models were included to study the impact of the noise on the performance of the algorithm (detailed in \cref{sec:imple}).

%%%%%%%%%%%%%%%%%%%%%%%%%%%%%%%%%%%%%%%%%%%%%%%%%%%%%%%%%%%%%%%%%%%%%%
% RESULTS / Energy estimate
%%%%%%%%%%%%%%%%%%%%%%%%%%%%%%%%%%%%%%%%%%%%%%%%%%%%%%%%%%%%%%%%%%%%%%
%%%%%%%%%%%%%%%%%%%%%%%%%%%%%%%%%%%%%%%%%%%%%%%%%%%%%%%%%%%%%%%%%%%%%%
%     File: Paper_energy.tex                               %
%     Tex Master: Paper.tex                               %
%                                                                    
%%%%%%%%%%%%%%%%%%%%%%%%%%%%%%%%%%%%%%%%%%%%%%%%%%%%%%%%%%%%%%%%%%%%%

\section{Estimation of the Energetic Cost}
\label{sec:resul}

The energy consumption of one implementation of the QFT algorithm was estimated theoretically. The methods that were used for this estimate were based on the approaches in \cite{main,asiani} and \cite{sagar}.

%%%%%%%%%%%%%%%%%%%%%%%%%%%%%%%%%%%%%%%%%%%%%%%%%%%%%%%%%%%%%%%%%%%%%%
\subsection{Energetic Costs of the Implementation of the QFT}

The optimized gate sequence that implements the QFT consists of a Hadamard gate, phased rotations, entangling gates and a swap gate.
The entangling gates are obtained from the free evolution of the system. Since there is no active manipulation of the ions, no energetic costs are incurred. The swap gate also does not involve energetic costs.

The theoretical energy estimate for the QFT gate sequence was obtained by calculating the energy carried by the microwave pulses applying each phased rotation gate,

\begin{equation}
    E_{\text{QFT}} = \sum_n \sum_{i \in G_n} E(\omega_n,\theta_i),
    \label{eq:energy-sum}
\end{equation}
where $n = 1,2,3$ denotes the qubits and $G_n$ is the set of phased rotations applied to each qubit along the sequence. $E(\omega_n,\theta_i)$ is the energy carried by the electromagnetic field of the microwave pulses, where $\omega_n$ is the resonance frequency of the addressed ion $n$ and $\theta_i$ is the rotation angle of the phased rotation $i$ in $G_n$. The energy is given by the expression 

\begin{equation}
    E(\omega,\theta) = P(\omega) \ \tau(\theta) = \frac{1}{2} \frac{A_{\text{rad}}(\omega)}{A_{\text{dip}}} \hbar \ \Omega \ \theta,
\end{equation}
where $A_{\text{rad}}(\omega)$ is the effective area for the microwave pulse wavefront, $A_{\text{dip}}$ is the effective dipole cross-section for the ion and $\Omega$ is the Rabi frequency (see appendix).

The DD pulses correspond to phased rotations with $(\theta = \pi$, $\phi = 0)$. The total cost of dynamical decoupling was estimated with

\begin{equation}
    E_{\text{DD}} = \sum_n \mathcal{N}_{\text{DD}} E(\omega_n,\pi),
\end{equation}
with $\mathcal{N}_{\text{DD}}=60$ representing the number of $\pi$ pulses addressed to each qubit during the algorithm.

The operations of preparation of the ions (Doppler cooling, sideband cooling, preparation in the ground state) and readout are only performed one time during the experiment, so the energetic costs were dubbed one-time costs, $E_{\text{one}}$, and were estimated with the expression

\begin{equation}
    E_{\text{one}} = \sum_{j=1}^4 E_{\text{one}}^j,
\end{equation}
where $j$ identifies the operation, which involves an energy consumption of $E_{\text{one}}^j$. 
The Doppler cooling ($j=1$) and the sideband cooling ($j=2$) are performed by using the $369.5$ nm laser, with powers $P^{369}_1=48$ $\mu$W, $P^{369}_2=0.16$ $\mu$W, and the microwave radiation, for time durations $t_1=8.0$ ms and $t_2=60$ ms, respectively. The energetic costs of cooling were determined by

\begin{equation}
    E_{\text{one}}^j = \Big( P^{369}_j + \sum_n P(\omega_n) \Big) t_j, \ \ j=1,2,
\end{equation}
where the sum is performed over the three qubits. 
The preparation in the ground state ($j=3$) and the readout ($j=4$) require only the $369.5$ nm laser with powers $P^{369}_3=35.0$ $\mu$W, $P^{369}_4=48.0$ $\mu$W, for time durations $t_3=0.20$ ms and $t_4=3.0$ ms. These steps involve an energy consumption of

\begin{equation}
    E_{\text{one}}^j = P^{369}_j  t_j, \ \ j=3,4.
    \label{eq:energy-one-34}
\end{equation}

A $935.2$ nm laser is continuously turned on during the whole experiment, from ion cooling to readout, with power $P^{935}=1.35$ mW, to maintain cooling and detection. The power necessary to maintain the Paul trap, $P_{\text{trap}}=10$ W, also contributes to the baseline costs. The power consumption of these two components occurs independently of what operation is being implemented on the quantum computer. 
The energetic cost of the baseline for the experiment was determined using the expression

\begin{equation}
    E_{\text{baseline}} = \big( P^{935} + P_{\text{trap}} \big) t_{\text{total}},
\end{equation}
where $t_{\text{total}} = t_{\text{QFT}} + t_{1-4} = 79.76$ ms is the total duration of the experiment, with $t_{\text{QFT}}=8.56$ ms and $t_{1-4}= t_1 + t_2 + t_3 + t_4 =71.20$ ms.

The power of the lasers corresponds to the optical power of the laser beams measured with a commercial optical power meter \cite{sagar}.

The total energetic costs for the QFT, one-time operations and continuous operations, are summarized in \cref{tab:energy-total}.

\begin{table}[H]
    \small
    \centering
    \caption{Energetic costs of one implementation of the QFT algorithm. The costs of the Laser (935.2 nm) and Paul trap are given once, accounting for the total duration of $79.76$ ms.}
    \resizebox{\columnwidth}{!}{
    \begin{tabular}{l c c}
    \hline
    Operation &  Duration (ms) & Energy ($\mu$J)\\
    \hline \hline
    QFT sequence & $8.56$ & $2.16 \times 10^1$\\
    \hline
    Dynamical decoupling & - & $5.22 \times 10^2$ \\
    \hline
    Doppler cooling & $8.0$ & $6.96 \times 10^3$ \\
    Sideband cooling & $60.0$ & $5.22 \times 10^4$ \\
    \hline
    Ground state preparation & $0.20$ & $7.0 \times 10^{-3}$\\
    \hline
    Readout & $3.0$ & $1.44 \times 10^{-1}$\\
    \hline
    Laser (935.2 nm) & $79.76$ & $1.08 \times 10^2$ \\
    \hline
    Paul trap & $79.76$ & $7.98 \times 10^5$ \\
    \hline \hline
    Total & $79.76$ & $8.58 \times 10^5$\\
    \hline
    \end{tabular}
    }
    \label{tab:energy-total}
\end{table}

The total energy consumption estimated for one implementation of the QFT algorithm is $E_{\text{total}}=8.58 \times 10^5$ $\mu$J. The dominant energetic costs are due to the cooling of the ions and the power supply of the Paul trap. A significant amount of energy is also necessary for the dynamical decoupling pulses. The energy used for the QFT gate sequence, however, is among the lowest values, possibly indicating an energy-effective computation of the unitary reversible part of the algorithm. 
The energy estimates demonstrate that the different components of the implementation of the QFT have important contributions to the total energy consumption. 

%%%%%%%%%%%%%%%%%%%%%%%%%%%%%%%%%%%%%%%%%%%%%%%%%%%%%%%%%%%%%%%%%%%%%%
\subsection{Scaling of the Energetic Costs}
\label{sec:scaling}

Scalability is a significant challenge in the development of quantum computers, in particular, due to the energy requirements. 
Based on the three-qubit experiment, hypothetical extrapolations were made about the energetic costs of the implementation of the QFT when the number of qubits scales up.

Assuming perfect pulses, noise-free entangling gates and all-to-all connectivity, the QFT gate sequence for $n$ qubits requires a total of $2(n-1)$ phase gates, $n$ Hadamard gates and $\mathcal{O}(n^2)$ $\pi$ pulses for coupling and uncoupling ions, hence the total number of gates scales as $n^2$ \cite{simplified}. 
The phase gates are implemented by a phase shift of the driving microwave field, rather than a physical modification of the qubit, so this type of gate does not incur an energetic cost. 
The total energy necessary to implement the QFT on $n$ qubits was thus estimated with the formula
\begin{equation}
    E_{\text{QFT}}(n) = n^2 E_{\pi} + n E_H,
\end{equation}
where $E_{\pi} = E (\omega, \pi)$ and $E_H = E\Big(\omega,\frac{\pi}{2}\Big) + E(\omega, \pi)$ are the energetic costs of the $\pi$ pulses and Hadamard gates, respectively.

The time duration of the $n-$dimensional QFT gate sequence grows with $n$ and is given by $T(n) \sim a n$ \cite{simplified}, with $a = t_{\text{QFT}}/3$. 
As the number of qubits scales up and the duration of the evolution times in the algorithm increases, more DD pulses are required to maintain the coherence of the qubits throughout the whole experiment. For the implementation considered, the coherence time increases approximately linearly with the number of DD pulses that are applied to each qubit \cite{khromova}. Because the DD sequences are applied to each of the $n$ qubits, the total number of pulses and the total energy required for dynamical decoupling increase with $n^2$. However, it should be noted that current ion traps have longer coherence times and shorter gate times, which will translate in a reduced number of required DD pulses.

During the preparation of the ions, the microwave radiation used for Doppler and sideband cooling is applied individually to each qubit, so the cooling energy is expected to increase with $n$, assuming that the $369.5$ nm laser that addresses the three qubits at the same time can still address a larger number of qubits.

Regarding the baseline, it was also assumed that the $935.2$ nm laser addressed all the qubits. However, the Paul trap can only hold up to a limited number of ions. For example, it is mentioned that traps with a string of $40$ ions are possible \cite{sizetrap}. 
The most straightforward manner of scaling the experimental setup is to consider placing multiple linear Paul traps holding $40$ ions end-to-end using precise electrode structures, although this architecture would limit the connectivity of the system \cite{trapped-ions-computing}.

\begin{figure}[ht]
    \centering
    \includegraphics[width=\columnwidth]{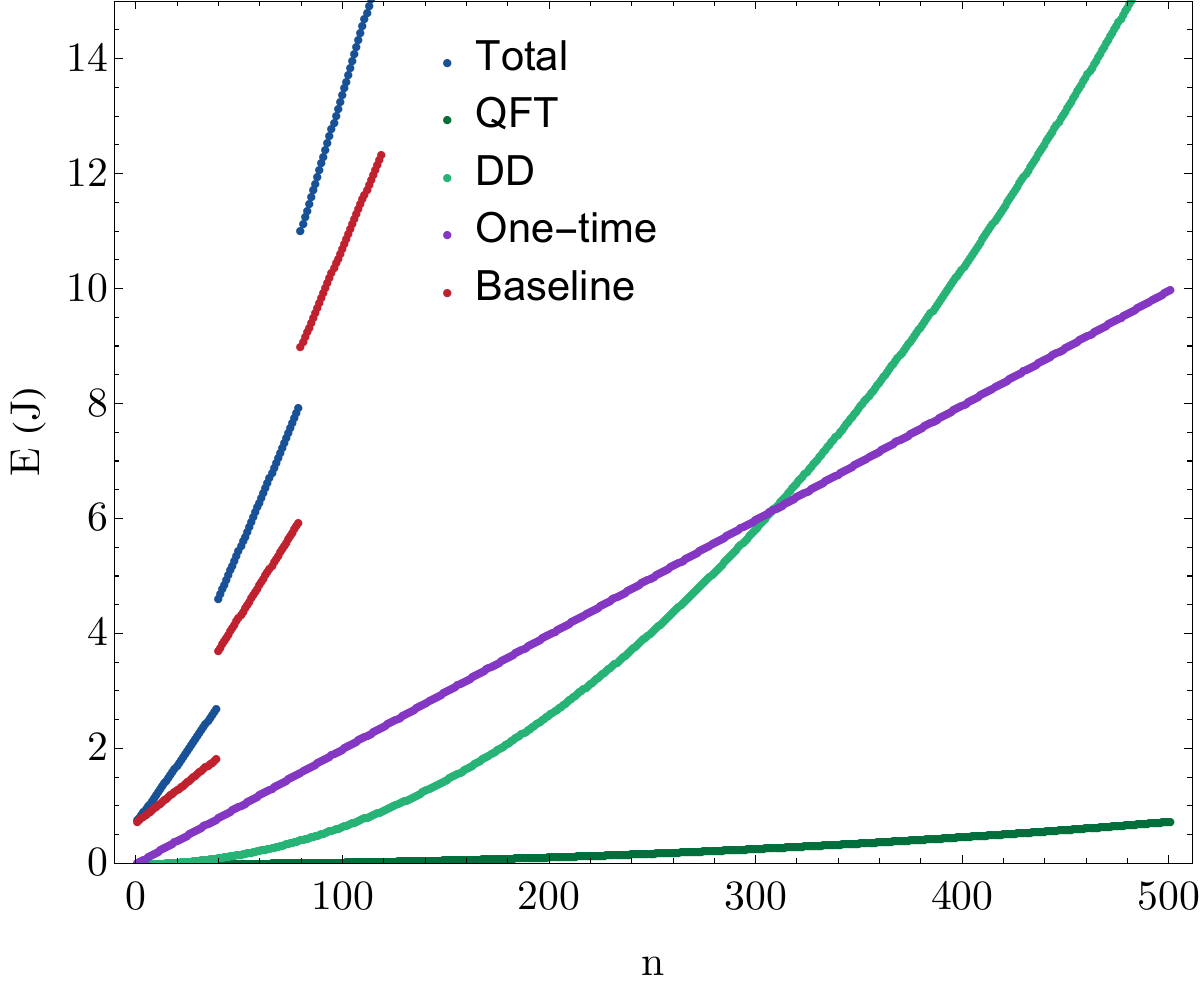}
    \caption{Scaling of the energetic costs of an implementation of the QFT with the number of qubits, $n$. Represented are the energetic costs of the QFT gate sequence, dynamical decoupling (DD) pulses, one-time operations performed for the preparation of the ions and readout, the baseline and the total. }
    \label{fig:scaling}
\end{figure}

The scaling of the different energetic costs of one implementation of the QFT with the number of qubits, $n$, is shown in \cref{fig:scaling}.

The energetic cost of the QFT gate sequence scales quadratically with the number of qubits. Even up to $500$ qubits, the energy to perform the QFT is significantly lower than the other costs, which points again to an energy-efficient computation. The energy used for dynamical decoupling also scales quadratically and it rapidly becomes one of the dominant energetic costs for large numbers of qubits. The one-time energetic costs of the preparation sequence and readout increase linearly with $n$. The most significant costs are due to the baseline. The jumps in the baseline curve are due to assuming that each trap can hold up to 40 ions, beyond which an additional trap is required. Although these considerations about scaling are extrapolations based on a number of assumptions, they convey that it is important to contemplate energetic concerns in the development of scalable quantum computers.

When scaling beyond a certain point, quantum error correction would become necessary to address the noise affecting the system. Furthermore, the algorithm considered here requires all-to-all connectivity, for which the shuttling of ions between the traps is needed. Both of these aspects require a decomposition of the QFT into sequences carried out for smaller groups of ions and a different analysis of the complexity in terms of the number of operations and required time, therefore impacting the energetic cost.

%%%%%%%%%%%%%%%%%%%%%%%%%%%%%%%%%%%%%%%%%%%%%%%%%%%%%%%%%%%%%%%%%%%%%%
\subsection{Towards an Energetic Quantum Advantage}

%{\color{red} Check everywhere for updating the classical computers to El Capitan and Kairos.}

To benchmark the estimated energetic efficiency of the QFT implemented on a quantum computer, the energetic costs were compared to the state-of-the-art of classical supercomputing devices performing the algorithm for the discrete Fourier transform (DFT), which is the classical analogue of the QFT \cite{preskill2024}. 

The performance of classical computers is typically measured in Flop/s (floating point operations per second). The Top 500 project \cite{top500} lists the most powerful classical computers in the World, ranking them by performance and computing efficiency in solving a dense system of linear equations. 
According to the latest ranking (November 2025), the fastest supercomputer in the world is El Capitan, with a maximal performance of $1809.00$ %$1195.00$
PetaFlop/s and power usage of $29685$ %$22786$
kW. 
Dividing the total power by the performance in Flops per second, we obtain the energetic cost of a single Flop, \cite{advantage}, in this case $1.64 \times 10^{-11}$ %$1.90 \times 10^{-14}$ 
J/Flop. 
The most energy-efficient supercomputer is Kairos, %Henri, 
performing at  $3.05$ %$2.88$
PetaFlop/s with an energy efficiency of $73.282$ %$65.396$ 
GFlops/W. The energy cost per Flop of Kairos %Henri
is then $1.365 \times 10^{-11}$ %$1.53 \times 10^{-14}$ 
J/Flop.
 
\begin{figure}[H]
    \includegraphics[width=\linewidth]{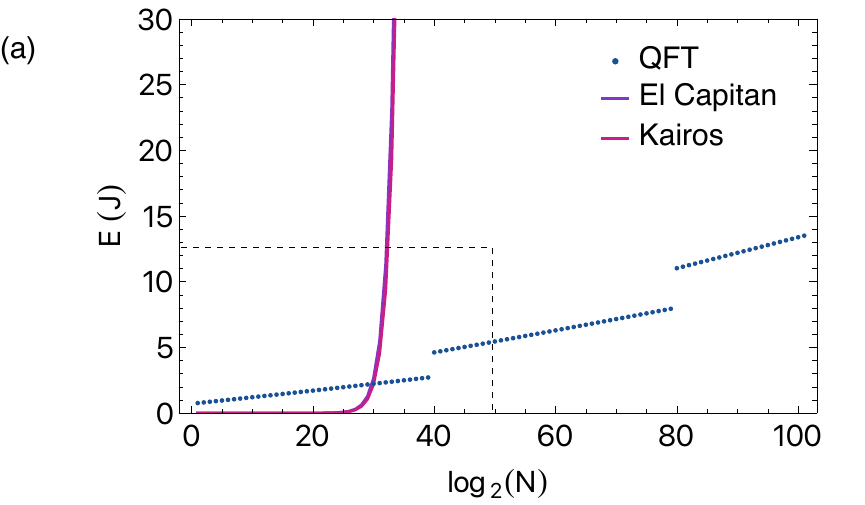}
     \includegraphics[width=\linewidth]{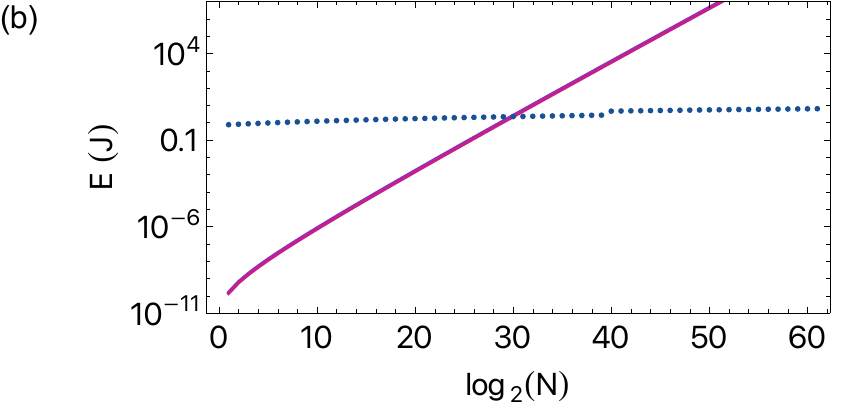}
    \includegraphics[width=\linewidth]{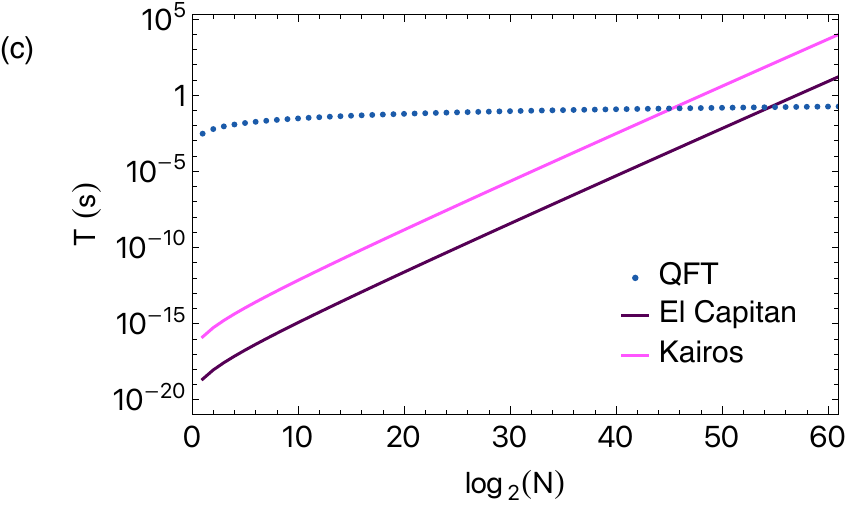}   
    \caption{(a) Scaling with the input size $N$ (represented through its logarithm) of the energetic costs of the implementation of the QFT on a trapped-ion quantum computer, compared to implementations of the discrete Fourier transform on the state-of-the-art classical supercomputers El Capitan and Kairos. For the quantum computer, an input of size $N$ requires $n=\log_2(N)$ qubits. (b) The section of the plot in the dashed box is shown with a logarithmic scale. (c) Scaling of the computational time.  Comparing panels (b) and (c) it is evident that the quantum energetic advantage is achieved for a smaller input size compared to the threshold for the computational time advantage. See the main text for more details.}
    \label{fig:qe-advantage}
\end{figure}

To compare the energy and time performances of the classical and quantum Fourier transform, we express their scaling as a function of the input size $N$. For the QFT, an input of size $N$ requires $n = \log_2(N)$ qubits to be encoded. On the classical front, a straight implementation of the DFT requires $\mathcal{O}(N^2)$ floating point operations; this, however, can be reduced to $\mathcal{O}(N \log_2(N))$ thanks to the Fast Fourier Transform (FFT) algorithm. Even in this case, the QFT scales exponentially faster than the FFT. The exact scaling of the QFT has been computed in Sec.~\ref{sec:scaling}, whereas for the FFT we rely on \cite{FFT_flops}, where it is reported that $5 N \log_2 (N)$ flops are required for an input of size $N$. The scaling behaviours of energy and time are obtained by multiplying the algorithmic scaling by the energy and time per flop, respectively.

\begin{comment}
\begin{figure}[h]
    \centering
    \begin{minipage}{\columnwidth}
    \centering
    \includegraphics[width=\linewidth]{NewEnergyAdvantage.pdf}
    \end{minipage}
    \begin{minipage}{\columnwidth}
    \centering
    \includegraphics[width=\linewidth]{NewEnergyAdvantageLog.pdf}
    \end{minipage}
    \begin{minipage}{\columnwidth}
    \centering
    \includegraphics[width=\linewidth]{NewTimeAdvantage.pdf}
    \end{minipage}
    \caption{(a) Scaling with the input size $N$ (represented through its logarithm) of the energetic costs of the implementation of the QFT on a trapped-ion quantum computer, compared to implementations of the discrete Fourier transform on the state-of-the-art classical supercomputers El Capitan and Kairos. For the quantum computer, an input of size $N$ requires $n=\log_2(N)$ qubits. (b) The section of the plot in the dashed box is shown with a logarithmic scale. (c) Scaling of the computational time. The gap between the two classical curves reflects the larger computational power of El Capitan with respect to Kairos. Comparing figures (b) and (c) it is evident that the quantum energetic advantage is achieved for a smaller input size.}
    \label{fig:qe-advantage}
\end{figure}
\end{comment}

%The gap between the two classical curves reflects the larger computational power of El Capitan with respect to Kairos.

\Cref{fig:qe-advantage} features a comparison of the scaling, as the input size $N$ increases, of the energy consumption of a QFT performed in a quantum computer with its classical analogue implemented on El Capitan and Kairos. 

For a small input size, the energetic cost of the implementation of the QFT is significantly higher than that of the classical Fourier transform. However, it was found that an energetic quantum advantage, compared to both supercomputers El Capitan and Kairos, may be achieved when the QFT is computed with an input size requiring at least 30 qubits, a number already within the capabilities of the most advanced current quantum processors. As for the time scalings, Fig.~\ref{fig:qe-advantage}, panel (c) shows that a quantum advantage can be expected from input size corresponding to $55$ qubits. Comparing panels (b) and (c) in Fig.~\ref{fig:qe-advantage}, we identify a range for the input size for which the quantum computation is more energy efficient than the classical, even though not yet faster. This can be understood comparing the energies required for a single quantum operation and for a flop, and the respective times. At the single operation level, a quantum gate requires both more energy and more time than a flop: comparing to El Capitan, the ratio [energy-per-gate]/[energy-per-flop] asymptotically is $4 \times 10^{9}$, whereas the ratio [time-per-gate]/[time-per-flop] tends to $2.5 \times 10^{16}$. Being the ratio more favourable for the energy, the exponential speedup of the QFT leads to the energetic quantum advantage appearing for a smaller input size than the time advantage.

These results were obtained in an ideal scenario, which does not include a noise model or quantum error correction. A more accurate comparison would consider the details of the specific classical implementation of the DFT, especially given the flexibility allowed by current systems which usually comprise a combination of CPUs and GPUs. Furthermore, in our comparison we studied the scaling behavior of single-shot realizations of both the QFT and the classical DFT; in particular, we considered the final readout, or measurement, as a one-time cost. Our analysis follows the experiment~\cite{versatile} which inspired the present work, where the reconstruction of the diagonal of the density matrix provided enough data for the certification of the correct execution of the QFT on selected input states, and for the assessment of its fidelity through the chosen metrics of performance (square statistical overlap and distinguishability). The results of our simulations, presented in Appendix~\ref{sec:simulation}, follow a similar approach. A more thorough comparison of the QFT and DFT with a generic input state would require performing a full tomographic reconstruction of the final quantum state, which requires a number of repetitions increasing exponentially with the number of qubits. In practice, however, the QFT finds its most useful applications not as a stand-alone algorithm, but as part of other algorithms, such as phase-estimation, order-finding and factoring.

Including the impact of quantum error correction on the scaling of the energetic performance will be a necessary step, as the number of qubits in quantum processors grows and realizations of error correcting codes become more common. As an example, in~\cite{stevens2026energyerrortradeoffencodingquantum} the authors find a tradeoff between required energy and computational precision in error-corrected quantum computation. In the case considered here, this would require designing and studying an error correcting implementation of the QFT with trapped ions, a task beyond the scope of the present work. It can be argued that error correction may cause at most a polynomial overhead, therefore the claim for an energetic quantum advantage should hold even in this scenario; a proper answer to this question, however, can only be obtained by studying real case studies of quantum error-correcting codes designed to be implemented on trapped-ion quantum processors, and the actual complexity will depend on the specific algorithm and implementation.

Another aspect to be considered in the scaling up of trapped ion quantum processors is the overhead due to utilizing multiple traps, as explained in Sec.~\ref{sec:scaling}. In order for a multi-trap setup to function as a single quantum processor, interactions between ions belonging to different traps must be implemented. A way to achieve this is, for example, shuttling ions belonging to different traps to a middle interaction region, where multi-qubit gates can be performed; this will require many additional operations, each with an associated energy cost. 
Finally, most likely future trapped-ion setups will require a cryogenic environment, to achieve a better vacuum, thus fewer background gas collisions, resulting in longer coherence times. It is also expected that a cryogenic environment will reduce the heating rate of trapped ion processors. Operating a quantum processor in a cryogenic environment comes at a significant energetic cost, most likely higher than any other cost considered so far, as it is also the case for superconducting circuits or quantum dots in semiconductors. This cost, however, is expected to be constant and independent of the number of qubits, at least until the internal size of the cryostat will pose a limitation on the size of the quantum processor. Furthermore, if the energetic cost of cryogenic cooling is to be accounted for, in view of achieving the most comprehensive estimation of the energetic requirements of trapped-ion quantum computation, it would be desirable to include also a more thorough estimation of the power requirements of generating laser pulses, microwave pulses, of the ion trap electromagnetic fields and of vacuum systems, although these costs are all expected to be both lower than those required for cryogenic cooling, and nearly constant the number of ions, at least up to a reasonable maximum number of ions.

Despite the approximations of the scaling model used, these observations can provide a new motivation for building quantum computers, towards the goal of building more energetically efficient technologies.

%%%%%%%%%%%%%%%%%%%%%%%%%%%%%%%%%%%%%%%%%%%%%%%%%%%%%%%%%%%%%%%%%%%%%%
% CONCLUSIONS
%%%%%%%%%%%%%%%%%%%%%%%%%%%%%%%%%%%%%%%%%%%%%%%%%%%%%%%%%%%%%%%%%%%%%%
%%%%%%%%%%%%%%%%%%%%%%%%%%%%%%%%%%%%%%%%%%%%%%%%%%%%%%%%%%%%%%%%%%%%%%
%     File: Paper_concl.tex                               %
%     Tex Master: Paper.tex                               %
%                                                                    
%%%%%%%%%%%%%%%%%%%%%%%%%%%%%%%%%%%%%%%%%%%%%%%%%%%%%%%%%%%%%%%%%%%%%

\section{Conclusions}
\label{sec:concl}

This work aimed at characterizing theoretically the energetics of an implementation of the quantum Fourier transform algorithm in a trapped-ion quantum processor. Our analysis was built upon data acquired through direct measurements in a similar trapped-ion setup,  presented in~\cite{sagar}.

The energetic cost of the implementation was estimated by describing the energy consumption of different elements of the setup and taking into account all the stages involved in the computation. 
The total energetic cost of the QFT gate sequence was found to be low, pointing towards an energy-efficient quantum information processing. In fact, the dominating energetic cost is found to be the supply of the Paul trap. %The estimate of the total energetic cost of the experiment includes important contributions from all its different components, establishing the need for a holistic approach in the characterization of the energy requirements of quantum computers. 

A possible scaling of the energetic costs of the QFT implementation with the number of qubits was studied. Some assumptions had to be taken about the scaling of some energetic costs, and further investigation is required; in spite of this, the comparison between the obtained scaling of the energetic costs of the QFT and state-of-the-art classical supercomputers performing the FFT showed that an energetic quantum advantage may be possible, with the quantum computer solving problems using less energy than a classical supercomputer. Remarkably, the energy advantage seems to be achievable for a rather small input size, already within the reach of the best available quantum computers. The comparison also reveals that the energetic advantage may be achievable for a smaller input size than required for the computational time advantage. The impact of this is twofold: first, this implies that, whenever trapped-ion quantum computers may offer a time advantage, they will also be energetically favourable, thus dispelling the concerns that quantum computing may be unfeasible from an energetic standpoint. Second, this reinforces the possibility that a quantum computer may be chosen over a classical one to perform a specific computational task not because of speed, but in virtue of energy convenience.

Further investigation towards optimizing the energetic costs will require identifying key control parameters of the experiment. In addition, the study of the scaling of the energetic costs can be refined by investigating different ion trap architectures, such as surface traps and error correction codes. Another possibility is the investigation of the scaling of the errors introduced by the quantum noise, and relating it to the energy consumption of the quantum computer when some error correction is included.
By addressing the energetic costs of the state preparation, computation and readout, and the baseline energy expenditure, this work thus provided the first steps of our agenda for a comprehensive study of the energetics of trapped-ion quantum computation. The impact of the classical control of the experiment, of ion transport and of quantum error correction will be  object of future work.

%%%%%%%%%%%%%%%%%%%%%%%%%%%%%%%%%%%%%%%%%%%%%%%%%%%%%%%%%%%%%%%%%%%%%%
% ACKNOWLEDGMENTS
%%%%%%%%%%%%%%%%%%%%%%%%%%%%%%%%%%%%%%%%%%%%%%%%%%%%%%%%%%%%%%%%%%%%%%
%%%%%%%%%%%%%%%%%%%%%%%%%%%%%%%%%%%%%%%%%%%%%%%%%%%%%%%%%%%%%%%%%%%%%%
%     File: Paper_ackno.tex                               %
%     Tex Master: Paper.tex                               %
%                                                                    
%%%%%%%%%%%%%%%%%%%%%%%%%%%%%%%%%%%%%%%%%%%%%%%%%%%%%%%%%%%%%%%%%%%%%

\begin{acknowledgements}
    The authors gratefully acknowledge helpful discussions with Christof Wunderlich, Dylan Lewis, Enrico Trombetti, Matteo Turco, Patrick Huber, and Sougato Bose. The authors also thank the support from FCT -- Funda\c{c}\~{a}o para a Ci\^{e}ncia e a Tecnologia (Portugal), namely through project UID/PRR/04540/2025 and contract LA/P/0095/2020.
\end{acknowledgements}

%%%%%%%%%%%%%%%%%%%%%%%%%%%%%%%%%%%%%%%%%%%%%%%%%%%%%%%%%%%%%%%%%%%%%%
% REFERENCES
%%%%%%%%%%%%%%%%%%%%%%%%%%%%%%%%%%%%%%%%%%%%%%%%%%%%%%%%%%%%%%%%%%%%%%

% External bibliography database file in the BibTeX format (Paper_ref_db.bib)
%\bibliographystyle{plainurl}
\bibliography{Paper}

\begin{thebibliography}{30}%
\makeatletter
\providecommand \@ifxundefined [1]{%
 \@ifx{#1\undefined}
}%
\providecommand \@ifnum [1]{%
 \ifnum #1\expandafter \@firstoftwo
 \else \expandafter \@secondoftwo
 \fi
}%
\providecommand \@ifx [1]{%
 \ifx #1\expandafter \@firstoftwo
 \else \expandafter \@secondoftwo
 \fi
}%
\providecommand \natexlab [1]{#1}%
\providecommand \enquote  [1]{``#1''}%
\providecommand \bibnamefont  [1]{#1}%
\providecommand \bibfnamefont [1]{#1}%
\providecommand \citenamefont [1]{#1}%
\providecommand \href@noop [0]{\@secondoftwo}%
\providecommand \href [0]{\begingroup \@sanitize@url \@href}%
\providecommand \@href[1]{\@@startlink{#1}\@@href}%
\providecommand \@@href[1]{\endgroup#1\@@endlink}%
\providecommand \@sanitize@url [0]{\catcode `\\12\catcode `\$12\catcode
  `\&12\catcode `\#12\catcode `\^12\catcode `\_12\catcode `\%12\relax}%
\providecommand \@@startlink[1]{}%
\providecommand \@@endlink[0]{}%
\providecommand \url  [0]{\begingroup\@sanitize@url \@url }%
\providecommand \@url [1]{\endgroup\@href {#1}{\urlprefix }}%
\providecommand \urlprefix  [0]{URL }%
\providecommand \Eprint [0]{\href }%
\providecommand \doibase [0]{https://doi.org/}%
\providecommand \selectlanguage [0]{\@gobble}%
\providecommand \bibinfo  [0]{\@secondoftwo}%
\providecommand \bibfield  [0]{\@secondoftwo}%
\providecommand \translation [1]{[#1]}%
\providecommand \BibitemOpen [0]{}%
\providecommand \bibitemStop [0]{}%
\providecommand \bibitemNoStop [0]{.\EOS\space}%
\providecommand \EOS [0]{\spacefactor3000\relax}%
\providecommand \BibitemShut  [1]{\csname bibitem#1\endcsname}%
\let\auto@bib@innerbib\@empty
%</preamble>
\bibitem [{\citenamefont {Abah}\ \emph {et~al.}(2019)\citenamefont {Abah},
  \citenamefont {Puebla}, \citenamefont {Kiely}, \citenamefont {De~Chiara},
  \citenamefont {Paternostro},\ and\ \citenamefont {Campbell}}]{en-control}%
  \BibitemOpen
  \bibfield  {author} {\bibinfo {author} {\bibfnamefont {O.}~\bibnamefont
  {Abah}}, \bibinfo {author} {\bibfnamefont {R.}~\bibnamefont {Puebla}},
  \bibinfo {author} {\bibfnamefont {A.}~\bibnamefont {Kiely}}, \bibinfo
  {author} {\bibfnamefont {G.}~\bibnamefont {De~Chiara}}, \bibinfo {author}
  {\bibfnamefont {M.}~\bibnamefont {Paternostro}},\ and\ \bibinfo {author}
  {\bibfnamefont {S.}~\bibnamefont {Campbell}},\ }\bibfield  {title} {\bibinfo
  {title} {Energetic cost of quantum control protocols},\ }\href
  {https://doi.org/10.1088/1367-2630/ab4c8c} {\bibfield  {journal} {\bibinfo
  {journal} {New Journal of Physics}\ }\textbf {\bibinfo {volume} {21}},\
  \bibinfo {pages} {103048} (\bibinfo {year} {2019})}\BibitemShut {NoStop}%
\bibitem [{\citenamefont {Guryanova}\ \emph {et~al.}(2020)\citenamefont
  {Guryanova}, \citenamefont {Friis},\ and\ \citenamefont {Huber}}]{en-meas}%
  \BibitemOpen
  \bibfield  {author} {\bibinfo {author} {\bibfnamefont {Y.}~\bibnamefont
  {Guryanova}}, \bibinfo {author} {\bibfnamefont {N.}~\bibnamefont {Friis}},\
  and\ \bibinfo {author} {\bibfnamefont {M.}~\bibnamefont {Huber}},\ }\bibfield
   {title} {\bibinfo {title} {Ideal projective measurements have infinite
  resource costs},\ }\href {https://doi.org/10.22331/q-2020-01-13-222}
  {\bibfield  {journal} {\bibinfo  {journal} {Quantum}\ }\textbf {\bibinfo
  {volume} {4}},\ \bibinfo {pages} {222} (\bibinfo {year} {2020})}\BibitemShut
  {NoStop}%
\bibitem [{\citenamefont {Gea-Banacloche}(2002)}]{en-min}%
  \BibitemOpen
  \bibfield  {author} {\bibinfo {author} {\bibfnamefont {J.}~\bibnamefont
  {Gea-Banacloche}},\ }\bibfield  {title} {\bibinfo {title} {Minimum energy
  requirements for quantum computation},\ }\href
  {https://doi.org/10.1103/PhysRevLett.89.217901} {\bibfield  {journal}
  {\bibinfo  {journal} {Physical Review Letters}\ }\textbf {\bibinfo {volume}
  {89}},\ \bibinfo {pages} {217901} (\bibinfo {year} {2002})}\BibitemShut
  {NoStop}%
\bibitem [{\citenamefont {Ikonen}\ \emph {et~al.}(2017)\citenamefont {Ikonen},
  \citenamefont {Salmilehto},\ and\ \citenamefont
  {M{\"o}tt{\"o}nen}}]{en-eneff}%
  \BibitemOpen
  \bibfield  {author} {\bibinfo {author} {\bibfnamefont {J.}~\bibnamefont
  {Ikonen}}, \bibinfo {author} {\bibfnamefont {J.}~\bibnamefont {Salmilehto}},\
  and\ \bibinfo {author} {\bibfnamefont {M.}~\bibnamefont {M{\"o}tt{\"o}nen}},\
  }\bibfield  {title} {\bibinfo {title} {Energy-efficient quantum computing},\
  }\href {https://doi.org/10.1038/s41534-017-0015-5} {\bibfield  {journal}
  {\bibinfo  {journal} {npj Quantum Information}\ }\textbf {\bibinfo {volume}
  {3}},\ \bibinfo {pages} {17} (\bibinfo {year} {2017})}\BibitemShut {NoStop}%
\bibitem [{\citenamefont {Stevens}\ \emph {et~al.}(2022)\citenamefont
  {Stevens}, \citenamefont {Szombati}, \citenamefont {Maffei}, \citenamefont
  {Elouard}, \citenamefont {Assouly}, \citenamefont {Cottet}, \citenamefont
  {Dassonneville}, \citenamefont {Ficheux}, \citenamefont {Zeppetzauer},
  \citenamefont {Bienfait}, \citenamefont {Jordan}, \citenamefont
  {Auff{\`{e}}ves},\ and\ \citenamefont {Huard}}]{en-sqgate}%
  \BibitemOpen
  \bibfield  {author} {\bibinfo {author} {\bibfnamefont {J.}~\bibnamefont
  {Stevens}}, \bibinfo {author} {\bibfnamefont {D.}~\bibnamefont {Szombati}},
  \bibinfo {author} {\bibfnamefont {M.}~\bibnamefont {Maffei}}, \bibinfo
  {author} {\bibfnamefont {C.}~\bibnamefont {Elouard}}, \bibinfo {author}
  {\bibfnamefont {R.}~\bibnamefont {Assouly}}, \bibinfo {author} {\bibfnamefont
  {N.}~\bibnamefont {Cottet}}, \bibinfo {author} {\bibfnamefont
  {R.}~\bibnamefont {Dassonneville}}, \bibinfo {author} {\bibfnamefont
  {Q.}~\bibnamefont {Ficheux}}, \bibinfo {author} {\bibfnamefont
  {S.}~\bibnamefont {Zeppetzauer}}, \bibinfo {author} {\bibfnamefont
  {A.}~\bibnamefont {Bienfait}}, \bibinfo {author} {\bibfnamefont
  {A.}~\bibnamefont {Jordan}}, \bibinfo {author} {\bibfnamefont
  {A.}~\bibnamefont {Auff{\`{e}}ves}},\ and\ \bibinfo {author} {\bibfnamefont
  {B.}~\bibnamefont {Huard}},\ }\bibfield  {title} {\bibinfo {title}
  {Energetics of a single qubit gate},\ }\href
  {https://doi.org/10.1103/PhysRevLett.129.110601} {\bibfield  {journal}
  {\bibinfo  {journal} {Physical Review Letters}\ }\textbf {\bibinfo {volume}
  {129}} (\bibinfo {year} {2022})}\BibitemShut {NoStop}%
\bibitem [{\citenamefont {Silva~Pratapsi}\ \emph {et~al.}(2023)\citenamefont
  {Silva~Pratapsi}, \citenamefont {Huber}, \citenamefont {Barthel},
  \citenamefont {Bose}, \citenamefont {Wunderlich},\ and\ \citenamefont
  {Omar}}]{sagar}%
  \BibitemOpen
  \bibfield  {author} {\bibinfo {author} {\bibfnamefont {S.}~\bibnamefont
  {Silva~Pratapsi}}, \bibinfo {author} {\bibfnamefont {P.~H.}\ \bibnamefont
  {Huber}}, \bibinfo {author} {\bibfnamefont {P.}~\bibnamefont {Barthel}},
  \bibinfo {author} {\bibfnamefont {S.}~\bibnamefont {Bose}}, \bibinfo {author}
  {\bibfnamefont {C.}~\bibnamefont {Wunderlich}},\ and\ \bibinfo {author}
  {\bibfnamefont {Y.}~\bibnamefont {Omar}},\ }\bibfield  {title} {\bibinfo
  {title} {Classical half-adder using trapped-ion quantum bits: Toward
  energy-efficient computation},\ }\href {https://doi.org/10.1063/5.0176719}
  {\bibfield  {journal} {\bibinfo  {journal} {Applied Physics Letters}\
  }\textbf {\bibinfo {volume} {123}},\ \bibinfo {pages} {154003} (\bibinfo
  {year} {2023})}\BibitemShut {NoStop}%
\bibitem [{\citenamefont {Moutinho}\ \emph {et~al.}(2023)\citenamefont
  {Moutinho}, \citenamefont {Pezzutto}, \citenamefont {Pratapsi}, \citenamefont
  {da~Silva}, \citenamefont {De~Franceschi}, \citenamefont {Bose},
  \citenamefont {Costa},\ and\ \citenamefont {Omar}}]{advantage}%
  \BibitemOpen
  \bibfield  {author} {\bibinfo {author} {\bibfnamefont {J.~P.}\ \bibnamefont
  {Moutinho}}, \bibinfo {author} {\bibfnamefont {M.}~\bibnamefont {Pezzutto}},
  \bibinfo {author} {\bibfnamefont {S.~S.}\ \bibnamefont {Pratapsi}}, \bibinfo
  {author} {\bibfnamefont {F.~F.}\ \bibnamefont {da~Silva}}, \bibinfo {author}
  {\bibfnamefont {S.}~\bibnamefont {De~Franceschi}}, \bibinfo {author}
  {\bibfnamefont {S.}~\bibnamefont {Bose}}, \bibinfo {author} {\bibfnamefont
  {A.~T.}\ \bibnamefont {Costa}},\ and\ \bibinfo {author} {\bibfnamefont
  {Y.}~\bibnamefont {Omar}},\ }\bibfield  {title} {\bibinfo {title} {Quantum
  dynamics for energetic advantage in a charge-based classical full adder},\
  }\href {https://doi.org/10.1103/PRXEnergy.2.033002} {\bibfield  {journal}
  {\bibinfo  {journal} {PRX Energy}\ }\textbf {\bibinfo {volume} {2}},\
  \bibinfo {pages} {033002} (\bibinfo {year} {2023})}\BibitemShut {NoStop}%
\bibitem [{\citenamefont {Lewis}\ \emph {et~al.}(2023)\citenamefont {Lewis},
  \citenamefont {Moutinho}, \citenamefont {Costa}, \citenamefont {Omar},\ and\
  \citenamefont {Bose}}]{dylan}%
  \BibitemOpen
  \bibfield  {author} {\bibinfo {author} {\bibfnamefont {D.}~\bibnamefont
  {Lewis}}, \bibinfo {author} {\bibfnamefont {J.~P.}\ \bibnamefont {Moutinho}},
  \bibinfo {author} {\bibfnamefont {A.~T.}\ \bibnamefont {Costa}}, \bibinfo
  {author} {\bibfnamefont {Y.}~\bibnamefont {Omar}},\ and\ \bibinfo {author}
  {\bibfnamefont {S.}~\bibnamefont {Bose}},\ }\bibfield  {title} {\bibinfo
  {title} {Low-dissipation data bus via coherent quantum dynamics},\ }\href
  {https://doi.org/10.1103/PhysRevB.108.075405} {\bibfield  {journal} {\bibinfo
   {journal} {Physical Review B}\ }\textbf {\bibinfo {volume} {108}},\ \bibinfo
  {pages} {075405} (\bibinfo {year} {2023})}\BibitemShut {NoStop}%
\bibitem [{\citenamefont {Martin}\ \emph {et~al.}(2022)\citenamefont {Martin},
  \citenamefont {Hughes}, \citenamefont {Moreno}, \citenamefont {Jones},
  \citenamefont {Sickinger}, \citenamefont {Narumanchi},\ and\ \citenamefont
  {Grout}}]{en-datacenters}%
  \BibitemOpen
  \bibfield  {author} {\bibinfo {author} {\bibfnamefont {M.~J.}\ \bibnamefont
  {Martin}}, \bibinfo {author} {\bibfnamefont {C.}~\bibnamefont {Hughes}},
  \bibinfo {author} {\bibfnamefont {G.}~\bibnamefont {Moreno}}, \bibinfo
  {author} {\bibfnamefont {E.~B.}\ \bibnamefont {Jones}}, \bibinfo {author}
  {\bibfnamefont {D.}~\bibnamefont {Sickinger}}, \bibinfo {author}
  {\bibfnamefont {S.}~\bibnamefont {Narumanchi}},\ and\ \bibinfo {author}
  {\bibfnamefont {R.}~\bibnamefont {Grout}},\ }\bibfield  {title} {\bibinfo
  {title} {Energy use in quantum data centers: Scaling the impact of computer
  architecture, qubit performance, size, and thermal parameters},\ }\href
  {https://doi.org/10.1109/TSUSC.2022.3190242} {\bibfield  {journal} {\bibinfo
  {journal} {IEEE Transactions on Sustainable Computing}\ }\textbf {\bibinfo
  {volume} {7}},\ \bibinfo {pages} {864} (\bibinfo {year} {2022})}\BibitemShut
  {NoStop}%
\bibitem [{\citenamefont {Landi}\ \emph {et~al.}(2020)\citenamefont {Landi},
  \citenamefont {de~Oliveira},\ and\ \citenamefont {Buksman}}]{en-error}%
  \BibitemOpen
  \bibfield  {author} {\bibinfo {author} {\bibfnamefont {G.~T.}\ \bibnamefont
  {Landi}}, \bibinfo {author} {\bibfnamefont {A.~L.~F.}\ \bibnamefont
  {de~Oliveira}},\ and\ \bibinfo {author} {\bibfnamefont {E.}~\bibnamefont
  {Buksman}},\ }\bibfield  {title} {\bibinfo {title} {Thermodynamic analysis of
  quantum error-correcting engines},\ }\href
  {https://doi.org/10.1103/PhysRevA.101.042106} {\bibfield  {journal} {\bibinfo
   {journal} {Physical Review A}\ }\textbf {\bibinfo {volume} {101}},\ \bibinfo
  {pages} {042106} (\bibinfo {year} {2020})}\BibitemShut {NoStop}%
\bibitem [{\citenamefont {Jaschke}\ and\ \citenamefont {Montangero}()}]{green}%
  \BibitemOpen
  \bibfield  {author} {\bibinfo {author} {\bibfnamefont {D.}~\bibnamefont
  {Jaschke}}\ and\ \bibinfo {author} {\bibfnamefont {S.}~\bibnamefont
  {Montangero}},\ }\href {https://doi.org/10.48550/arXiv.2205.12092} {\bibinfo
  {title} {Is quantum computing green? {A}n estimate for an energy-efficiency
  quantum advantage}},\ \bibinfo {note} {arXiv:2205.12092 (2022)}\BibitemShut
  {NoStop}%
\bibitem [{\citenamefont {Fellous-Asiani}\ \emph {et~al.}(2023)\citenamefont
  {Fellous-Asiani}, \citenamefont {Chai}, \citenamefont {Thonnart},
  \citenamefont {Ng}, \citenamefont {Whitney},\ and\ \citenamefont
  {Auff{\`e}ves}}]{main}%
  \BibitemOpen
  \bibfield  {author} {\bibinfo {author} {\bibfnamefont {M.}~\bibnamefont
  {Fellous-Asiani}}, \bibinfo {author} {\bibfnamefont {J.~H.}\ \bibnamefont
  {Chai}}, \bibinfo {author} {\bibfnamefont {Y.}~\bibnamefont {Thonnart}},
  \bibinfo {author} {\bibfnamefont {H.~K.}\ \bibnamefont {Ng}}, \bibinfo
  {author} {\bibfnamefont {R.~S.}\ \bibnamefont {Whitney}},\ and\ \bibinfo
  {author} {\bibfnamefont {A.}~\bibnamefont {Auff{\`e}ves}},\ }\bibfield
  {title} {\bibinfo {title} {Optimizing resource efficiencies for scalable
  full-stack quantum computers},\ }\href
  {https://doi.org/10.1103/PRXQuantum.4.040319} {\bibfield  {journal} {\bibinfo
   {journal} {PRX Quantum}\ }\textbf {\bibinfo {volume} {4}},\ \bibinfo {pages}
  {040319} (\bibinfo {year} {2023})}\BibitemShut {NoStop}%
\bibitem [{\citenamefont {Asiani}()}]{asiani}%
  \BibitemOpen
  \bibfield  {author} {\bibinfo {author} {\bibfnamefont {M.~F.}\ \bibnamefont
  {Asiani}},\ }\emph {\bibinfo {title} {The resource cost of large scale
  quantum computing}},\ \href {https://doi.org/10.48550/arXiv.2112.04022}
  {Ph.D. thesis},\ \bibinfo  {school} {Universit{\'e} Grenoble Alpes},\
  \bibinfo {note} {arXiv:2112.04022 (2021)}\BibitemShut {NoStop}%
\bibitem [{\citenamefont {Stevens}\ and\ \citenamefont
  {Deffner}(2025)}]{Stevens_2025}%
  \BibitemOpen
  \bibfield  {author} {\bibinfo {author} {\bibfnamefont {J.}~\bibnamefont
  {Stevens}}\ and\ \bibinfo {author} {\bibfnamefont {S.}~\bibnamefont
  {Deffner}},\ }\bibfield  {title} {\bibinfo {title} {Hamiltonian quantum
  gates-energetic advantage from entangleability},\ }\href
  {https://doi.org/10.1088/2058-9565/ae0daf} {\bibfield  {journal} {\bibinfo
  {journal} {Quantum Science and Technology}\ }\textbf {\bibinfo {volume}
  {10}},\ \bibinfo {pages} {04LT03} (\bibinfo {year} {2025})}\BibitemShut
  {NoStop}%
\bibitem [{\citenamefont {Carrasco-Codina}\ \emph {et~al.}(2026)\citenamefont
  {Carrasco-Codina}, \citenamefont {Escofet}, \citenamefont {Hilaire},
  \citenamefont {Soret}, \citenamefont {Nerenberg}, \citenamefont {Champain},
  \citenamefont {Milburn}, \citenamefont {Li}, \citenamefont {Bautista},
  \citenamefont {Gómez}, \citenamefont {Miralles}, \citenamefont {Abadal},
  \citenamefont {Almudéver}, \citenamefont {G.}, \citenamefont {Alarcón},\
  and\ \citenamefont {Yehia}}]{Ariane2026}%
  \BibitemOpen
  \bibfield  {author} {\bibinfo {author} {\bibfnamefont {M.}~\bibnamefont
  {Carrasco-Codina}}, \bibinfo {author} {\bibfnamefont {P.}~\bibnamefont
  {Escofet}}, \bibinfo {author} {\bibfnamefont {P.}~\bibnamefont {Hilaire}},
  \bibinfo {author} {\bibfnamefont {A.}~\bibnamefont {Soret}}, \bibinfo
  {author} {\bibfnamefont {S.}~\bibnamefont {Nerenberg}}, \bibinfo {author}
  {\bibfnamefont {V.}~\bibnamefont {Champain}}, \bibinfo {author}
  {\bibfnamefont {K.}~\bibnamefont {Milburn}, \bibfnamefont
  {Gerard~Theophilo}}, \bibinfo {author} {\bibfnamefont {S.~H.}\ \bibnamefont
  {Li}}, \bibinfo {author} {\bibfnamefont {I.}~\bibnamefont {Bautista}},
  \bibinfo {author} {\bibfnamefont {A.}~\bibnamefont {Gómez}}, \bibinfo
  {author} {\bibfnamefont {J.}~\bibnamefont {Miralles}}, \bibinfo {author}
  {\bibfnamefont {S.}~\bibnamefont {Abadal}}, \bibinfo {author} {\bibnamefont
  {Almudéver}}, \bibinfo {author} {\bibfnamefont {C.}~\bibnamefont {G.}},
  \bibinfo {author} {\bibfnamefont {E.}~\bibnamefont {Alarcón}},\ and\
  \bibinfo {author} {\bibfnamefont {R.}~\bibnamefont {Yehia}},\ }\href
  {https://doi.org/10.48550/arXiv.2605.15090} {\bibinfo {title} {Energy
  efficiency of quantum computers}} (\bibinfo {year} {2026}),\ \Eprint
  {https://arxiv.org/abs/2605.15090} {arXiv:2605.15090 [quant-ph]} \BibitemShut
  {NoStop}%
\bibitem [{\citenamefont {Piltz}\ \emph {et~al.}(2016)\citenamefont {Piltz},
  \citenamefont {Sriarunothai}, \citenamefont {Ivanov}, \citenamefont {Wölk},\
  and\ \citenamefont {Wunderlich}}]{versatile}%
  \BibitemOpen
  \bibfield  {author} {\bibinfo {author} {\bibfnamefont {C.}~\bibnamefont
  {Piltz}}, \bibinfo {author} {\bibfnamefont {T.}~\bibnamefont {Sriarunothai}},
  \bibinfo {author} {\bibfnamefont {S.~S.}\ \bibnamefont {Ivanov}}, \bibinfo
  {author} {\bibfnamefont {S.}~\bibnamefont {Wölk}},\ and\ \bibinfo {author}
  {\bibfnamefont {C.}~\bibnamefont {Wunderlich}},\ }\bibfield  {title}
  {\bibinfo {title} {Versatile microwave-driven trapped ion spin system for
  quantum information processing},\ }\href
  {https://doi.org/10.1126/sciadv.1600093} {\bibfield  {journal} {\bibinfo
  {journal} {Science Advances}\ }\textbf {\bibinfo {volume} {2}},\ \bibinfo
  {pages} {e1600093} (\bibinfo {year} {2016})}\BibitemShut {NoStop}%
\bibitem [{\citenamefont {Alves}\ \emph {et~al.}(2026)\citenamefont {Alves},
  \citenamefont {Pezzutto},\ and\ \citenamefont {Omar}}]{Oscar2026}%
  \BibitemOpen
  \bibfield  {author} {\bibinfo {author} {\bibfnamefont {O.}~\bibnamefont
  {Alves}}, \bibinfo {author} {\bibfnamefont {M.}~\bibnamefont {Pezzutto}},\
  and\ \bibinfo {author} {\bibfnamefont {Y.}~\bibnamefont {Omar}},\ }\bibfield
  {title} {\bibinfo {title} {{Energetics of Rydberg-atom quantum computing}},\
  }\href {https://doi.org/10.48550/arXiv.2601.03141} {\bibfield  {journal}
  {\bibinfo  {journal} {arXiv preprint arXiv:2601.03141}\ } (\bibinfo {year}
  {2026})}\BibitemShut {NoStop}%
\bibitem [{\citenamefont {Santos}\ \emph {et~al.}(2026)\citenamefont {Santos},
  \citenamefont {Nath}, \citenamefont {Pezzutto}, \citenamefont {Meunier},\
  and\ \citenamefont {Omar}}]{Joao2026}%
  \BibitemOpen
  \bibfield  {author} {\bibinfo {author} {\bibfnamefont {J.}~\bibnamefont
  {Santos}}, \bibinfo {author} {\bibfnamefont {J.}~\bibnamefont {Nath}},
  \bibinfo {author} {\bibfnamefont {M.}~\bibnamefont {Pezzutto}}, \bibinfo
  {author} {\bibfnamefont {T.}~\bibnamefont {Meunier}},\ and\ \bibinfo {author}
  {\bibfnamefont {Y.}~\bibnamefont {Omar}},\ }\bibfield  {title} {\bibinfo
  {title} {Unveiling energetic advantage in spin qubits quantum computation},\
  }\href@noop {} {\bibfield  {journal} {\bibinfo  {journal} {arXiv preprint, in
  preparation}\ } (\bibinfo {year} {2026})}\BibitemShut {NoStop}%
\bibitem [{\citenamefont {Ramos}\ \emph {et~al.}(2026)\citenamefont {Ramos},
  \citenamefont {Pezzutto}, \citenamefont {Meunier},\ and\ \citenamefont
  {Omar}}]{Pedro2026}%
  \BibitemOpen
  \bibfield  {author} {\bibinfo {author} {\bibfnamefont {P.}~\bibnamefont
  {Ramos}}, \bibinfo {author} {\bibfnamefont {M.}~\bibnamefont {Pezzutto}},
  \bibinfo {author} {\bibfnamefont {T.}~\bibnamefont {Meunier}},\ and\ \bibinfo
  {author} {\bibfnamefont {Y.}~\bibnamefont {Omar}},\ }\bibfield  {title}
  {\bibinfo {title} {Unveiling energetic advantage in superconducting
  cat-qubits quantum computation},\ }\href@noop {} {\bibfield  {journal}
  {\bibinfo  {journal} {arXiv preprint, in preparation}\ } (\bibinfo {year}
  {2026})}\BibitemShut {NoStop}%
\bibitem [{\citenamefont {Mintert}\ and\ \citenamefont
  {Wunderlich}(2001{\natexlab{a}})}]{magic}%
  \BibitemOpen
  \bibfield  {author} {\bibinfo {author} {\bibfnamefont {F.}~\bibnamefont
  {Mintert}}\ and\ \bibinfo {author} {\bibfnamefont {C.}~\bibnamefont
  {Wunderlich}},\ }\bibfield  {title} {\bibinfo {title} {Ion-trap quantum logic
  using long-wavelength radiation},\ }\href
  {https://doi.org/10.1103/PhysRevLett.87.257904} {\bibfield  {journal}
  {\bibinfo  {journal} {Phys. Rev. Lett.}\ }\textbf {\bibinfo {volume} {87}},\
  \bibinfo {pages} {257904} (\bibinfo {year} {2001}{\natexlab{a}})}\BibitemShut
  {NoStop}%
\bibitem [{\citenamefont {Khromova}\ \emph {et~al.}(2012)\citenamefont
  {Khromova}, \citenamefont {Piltz}, \citenamefont {Scharfenberger},
  \citenamefont {Gloger}, \citenamefont {Johanning}, \citenamefont {Var\'on},\
  and\ \citenamefont {Wunderlich}}]{khromova}%
  \BibitemOpen
  \bibfield  {author} {\bibinfo {author} {\bibfnamefont {A.}~\bibnamefont
  {Khromova}}, \bibinfo {author} {\bibfnamefont {C.}~\bibnamefont {Piltz}},
  \bibinfo {author} {\bibfnamefont {B.}~\bibnamefont {Scharfenberger}},
  \bibinfo {author} {\bibfnamefont {T.~F.}\ \bibnamefont {Gloger}}, \bibinfo
  {author} {\bibfnamefont {M.}~\bibnamefont {Johanning}}, \bibinfo {author}
  {\bibfnamefont {A.~F.}\ \bibnamefont {Var\'on}},\ and\ \bibinfo {author}
  {\bibfnamefont {C.}~\bibnamefont {Wunderlich}},\ }\bibfield  {title}
  {\bibinfo {title} {Designer spin pseudomolecule implemented with trapped ions
  in a magnetic gradient},\ }\href
  {https://doi.org/10.1103/PhysRevLett.108.220502} {\bibfield  {journal}
  {\bibinfo  {journal} {Phys. Rev. Lett.}\ }\textbf {\bibinfo {volume} {108}},\
  \bibinfo {pages} {220502} (\bibinfo {year} {2012})}\BibitemShut {NoStop}%
\bibitem [{\citenamefont {Ivanov}\ \emph {et~al.}(2015)\citenamefont {Ivanov},
  \citenamefont {Johanning},\ and\ \citenamefont {Wunderlich}}]{simplified}%
  \BibitemOpen
  \bibfield  {author} {\bibinfo {author} {\bibfnamefont {S.~S.}\ \bibnamefont
  {Ivanov}}, \bibinfo {author} {\bibfnamefont {M.}~\bibnamefont {Johanning}},\
  and\ \bibinfo {author} {\bibfnamefont {C.}~\bibnamefont {Wunderlich}},\
  }\bibfield  {title} {\bibinfo {title} {Simplified implementation of the
  quantum fourier transform with ising-type hamiltonians: Example with ion
  traps},\ }\href {https://doi.org/10.48550/arXiv.1503.08806} {\bibfield
  {journal} {\bibinfo  {journal} {arXiv preprint arXiv:1503.08806}\ } (\bibinfo
  {year} {2015})}\BibitemShut {NoStop}%
\bibitem [{\citenamefont {Khromova}(2012)}]{phdthesis}%
  \BibitemOpen
  \bibfield  {author} {\bibinfo {author} {\bibfnamefont {A.}~\bibnamefont
  {Khromova}},\ }\emph {\bibinfo {title} {Quantum gates with trapped ions using
  magnetic gradient induced coupling}},\ \href@noop {} {Ph.D. thesis} (\bibinfo
  {year} {2012})\BibitemShut {NoStop}%
\bibitem [{\citenamefont {Mintert}\ and\ \citenamefont
  {Wunderlich}(2001{\natexlab{b}})}]{sizetrap}%
  \BibitemOpen
  \bibfield  {author} {\bibinfo {author} {\bibfnamefont {F.}~\bibnamefont
  {Mintert}}\ and\ \bibinfo {author} {\bibfnamefont {C.}~\bibnamefont
  {Wunderlich}},\ }\bibfield  {title} {\bibinfo {title} {Ion-trap quantum logic
  using long-wavelength radiation},\ }\href
  {https://doi.org/10.1103/PhysRevLett.87.257904} {\bibfield  {journal}
  {\bibinfo  {journal} {Phys. Rev. Lett.}\ }\textbf {\bibinfo {volume} {87}},\
  \bibinfo {pages} {257904} (\bibinfo {year} {2001}{\natexlab{b}})}\BibitemShut
  {NoStop}%
\bibitem [{\citenamefont {Bruzewicz}\ \emph {et~al.}(2019)\citenamefont
  {Bruzewicz}, \citenamefont {Chiaverini}, \citenamefont {McConnell},\ and\
  \citenamefont {Sage}}]{trapped-ions-computing}%
  \BibitemOpen
  \bibfield  {author} {\bibinfo {author} {\bibfnamefont {C.~D.}\ \bibnamefont
  {Bruzewicz}}, \bibinfo {author} {\bibfnamefont {J.}~\bibnamefont
  {Chiaverini}}, \bibinfo {author} {\bibfnamefont {R.}~\bibnamefont
  {McConnell}},\ and\ \bibinfo {author} {\bibfnamefont {J.~M.}\ \bibnamefont
  {Sage}},\ }\bibfield  {title} {\bibinfo {title} {Trapped-ion quantum
  computing: Progress and challenges},\ }\href
  {https://doi.org/10.1063/1.5088164} {\bibfield  {journal} {\bibinfo
  {journal} {Applied Physics Reviews}\ }\textbf {\bibinfo {volume} {6}}
  (\bibinfo {year} {2019})}\BibitemShut {NoStop}%
\bibitem [{\citenamefont {Preskill}(2024)}]{preskill2024}%
  \BibitemOpen
  \bibfield  {author} {\bibinfo {author} {\bibfnamefont {J.}~\bibnamefont
  {Preskill}},\ }\href
  {http://theory.caltech.edu/~preskill/ph219/ph219_2024.html} {\bibinfo {title}
  {Lecture notes for physics 219: Quantum computation}},\ \bibinfo
  {howpublished} {California Institute of Technology} (\bibinfo {year}
  {2024})\BibitemShut {NoStop}%
\bibitem [{top()}]{top500}%
  \BibitemOpen
  \href@noop {} {\bibinfo {title} {Top 500 project}},\ \bibinfo {howpublished}
  {\url{https://www.top500.org/project/}},\ \bibinfo {note} {accessed: April
  2024}\BibitemShut {NoStop}%
\bibitem [{\citenamefont {Van~Loan}(1992)}]{FFT_flops}%
  \BibitemOpen
  \bibfield  {author} {\bibinfo {author} {\bibfnamefont {C.}~\bibnamefont
  {Van~Loan}},\ }\href@noop {} {\emph {\bibinfo {title} {Computational
  Frameworks for the Fast Fourier Transform}}}\ (\bibinfo  {publisher} {Society
  for Industrial and Applied Mathematics},\ \bibinfo {year} {1992})\
  Chap.~\bibinfo {chapter} {1}, p.~\bibinfo {pages} {15},\ \bibinfo {note}
  {\href{https://doi.org/10.1137/1.9781611970999.ch1}{doi:
  10.1137/1.9781611970999.ch1}}\BibitemShut {NoStop}%
\bibitem [{\citenamefont {Stevens}\ and\ \citenamefont
  {Deffner}(2026)}]{stevens2026energyerrortradeoffencodingquantum}%
  \BibitemOpen
  \bibfield  {author} {\bibinfo {author} {\bibfnamefont {J.}~\bibnamefont
  {Stevens}}\ and\ \bibinfo {author} {\bibfnamefont {S.}~\bibnamefont
  {Deffner}},\ }\href {https://arxiv.org/abs/2605.04329} {\bibinfo {title}
  {Energy-error tradeoff in encoding quantum error correction}} (\bibinfo
  {year} {2026}),\ \Eprint {https://arxiv.org/abs/2605.04329} {arXiv:2605.04329
  [quant-ph]} \BibitemShut {NoStop}%
\bibitem [{\citenamefont {Majumder}\ \emph {et~al.}(2023)\citenamefont
  {Majumder}, \citenamefont {Yale}, \citenamefont {Morris}, \citenamefont
  {Lobser}, \citenamefont {Burch}, \citenamefont {Chow}, \citenamefont
  {Revelle}, \citenamefont {Clark},\ and\ \citenamefont {Pooser}}]{noisygate}%
  \BibitemOpen
  \bibfield  {author} {\bibinfo {author} {\bibfnamefont {S.}~\bibnamefont
  {Majumder}}, \bibinfo {author} {\bibfnamefont {C.~G.}\ \bibnamefont {Yale}},
  \bibinfo {author} {\bibfnamefont {T.~D.}\ \bibnamefont {Morris}}, \bibinfo
  {author} {\bibfnamefont {D.~S.}\ \bibnamefont {Lobser}}, \bibinfo {author}
  {\bibfnamefont {A.~D.}\ \bibnamefont {Burch}}, \bibinfo {author}
  {\bibfnamefont {M.~N.}\ \bibnamefont {Chow}}, \bibinfo {author}
  {\bibfnamefont {M.~C.}\ \bibnamefont {Revelle}}, \bibinfo {author}
  {\bibfnamefont {S.~M.}\ \bibnamefont {Clark}},\ and\ \bibinfo {author}
  {\bibfnamefont {R.~C.}\ \bibnamefont {Pooser}},\ }\bibfield  {title}
  {\bibinfo {title} {Characterizing and mitigating coherent errors in a trapped
  ion quantum processor using hidden inverses},\ }\href
  {https://doi.org/10.22331/q-2023-05-15-1006} {\bibfield  {journal} {\bibinfo
  {journal} {Quantum}\ }\textbf {\bibinfo {volume} {7}},\ \bibinfo {pages}
  {1006} (\bibinfo {year} {2023})}\BibitemShut {NoStop}%
\end{thebibliography}%

%%%%%%%
% APPENDIX
%%%%%%%%%%%%%%%%%%%%%%%%%%%%%%%%%%%%%%%%%%%%%%%%%%%%%%%%%%%%%%%%%%%%%%
\appendix

%%%%%%%%%%%%%%%%%%%%%%%%%%%%%%%%%%%%%%%%%%%%%%%%%%%%%%%%%%%%%%%%%%%%%%
%     File: Paper_simul.tex                               %
%     Tex Master: Paper.tex                               %
%                                                                    
%%%%%%%%%%%%%%%%%%%%%%%%%%%%%%%%%%%%%%%%%%%%%%%%%%%%%%%%%%%%%%%%%%%%%

\section{Quantum Fourier Transform Simulation}
\label{sec:imple}

To validate the gate sequence \eqref{eq:Uopt} in performing a QFT, a simulation of the circuit was created in \textit{Mathematica}.

%%%%%%%%%%%%%%%%%%%%%%%%%%%%%%%%%%%%%%%%%%%%%%%%%%%%%%%%%%%%%%%%%%%%%%
\subsection{Circuit simulation}
\label{sec:simulation}

The operation that realizes the QFT, applied to a given input state $\ket{\psi}$, is given by

\begin{equation}
    \ket{\psi} \mapsto \textit{SWAP}_{13} \cdot U_{\text{QFT}}^{\text{opt}} \ket{\psi}.
\end{equation}

This operation was simulated numerically by applying the gate sequence $\textit{SWAP}_{13} \cdot U_{\text{QFT}}^{\text{opt}}$ to the input states $\ket{\psi} \in$ $\{\ket{111},$ $\ket{+11},$ $\ket{++1},$ $\ket{+++}\}$, represented by the respective state vectors. The output state was given in terms of the probabilities of obtaining each of the computational basis states in a measurement, $p_i$.

The expected theoretical results of the QFT for the different input states were computed with \cref{eq:qft,eq:qfteqsup}. These results were also expressed in terms of the probabilities of measuring each of the computational basis states, $q_i$.

The results were compared qualitatively with probability plots, superimposing the histograms of both the simulated and the theoretical results. To compare quantitatively the simulated probabilities $p_i$ with the theoretical probabilities $q_i$, where $i=0,1,...,7$ are the labels for the computational basis states, $\{ \ket{000}$, $\ket{001}$, $\ket{010}$, $\ket{011}$, $\ket{100}$, $\ket{101}$, $\ket{110}$, $\ket{111}\}$, the following metrics of performance were used:

\begin{itemize}
    \item Squared statistical overlap:
    \begin{equation}
        \gamma_{\text{SSO}} (p,q) = \Big( \sum_i \sqrt{p_i q_i} \Big)^2
        \label{eq:sso}
    \end{equation}

    \item Distinguishability: 
    \begin{equation}
        D (p,q) = 1 - \frac{1}{2} \sum_i \abs{p_i - q_i}
        \label{eq:distinguishability}
    \end{equation}
\end{itemize}

\begin{figure}[H]
    \centering
    \begin{minipage}{0.45\columnwidth}
        \centering
        \includegraphics[width=\linewidth]{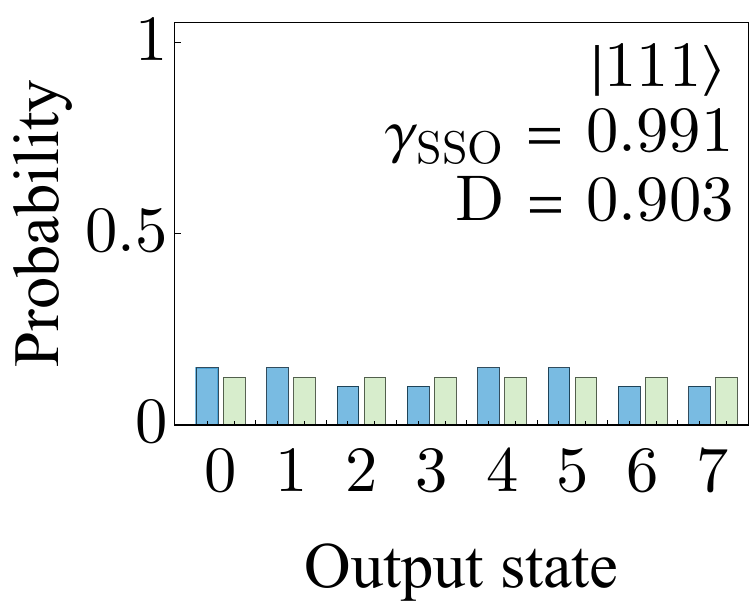}
    \end{minipage}
    \begin{minipage}{0.45\columnwidth}
        \centering
        \includegraphics[width=\linewidth]{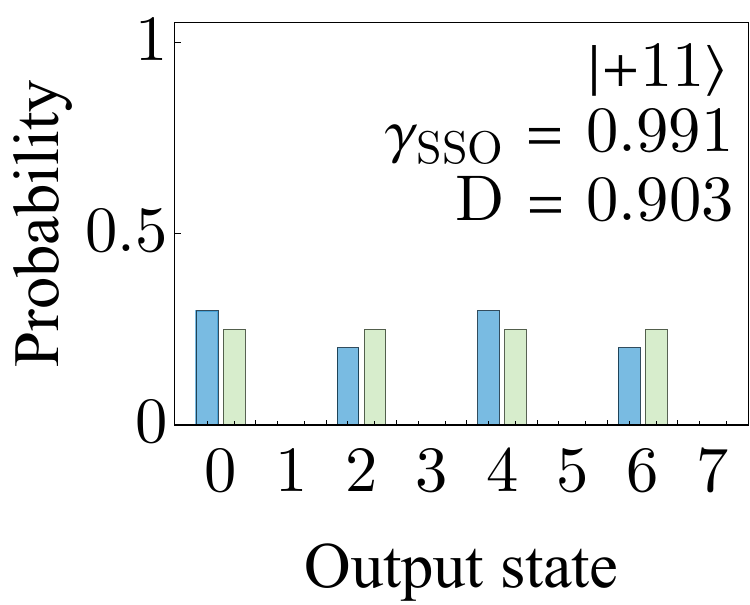}
    \end{minipage}
    \begin{minipage}{0.45\columnwidth}
        \centering
        \includegraphics[width=\linewidth]{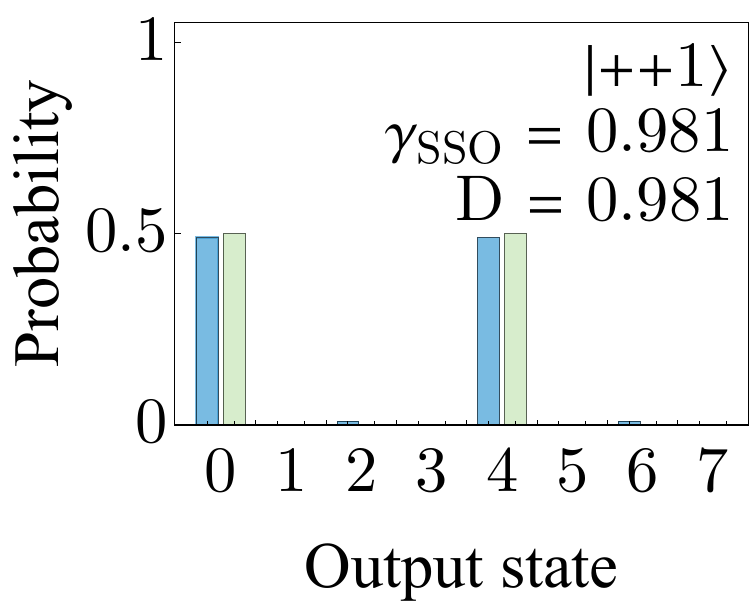}
    \end{minipage}
    \begin{minipage}{0.45\columnwidth}
        \centering
        \includegraphics[width=\linewidth]{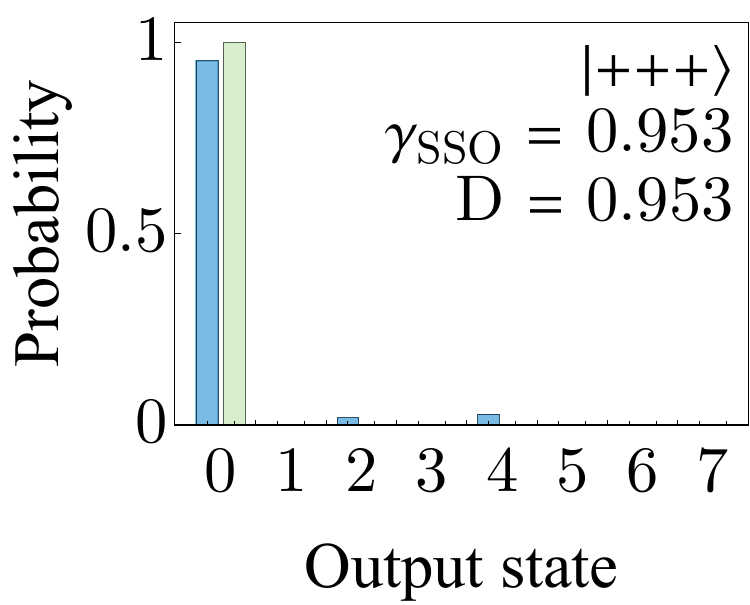}
    \end{minipage}
    \begin{minipage}{0.25\textwidth}
        \centering
        \includegraphics[width=\linewidth]{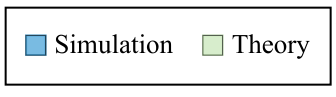}
    \end{minipage}
    \caption{Results of the QFT simulation, for the four input states $\{\ket{111},$ $\ket{+11},$ $\ket{++1},$ $\ket{+++}\}$, showing the probability of each output state occurring in a measurement. The simulation results (blue) are compared with the theoretically expected results (green). The SSO $\gamma_{\text{SSO}}$ and distinguishability $D$ are metrics of the performance.}
    \label{fig:simulation-4states}
\end{figure}

The QFT of input states $\{\ket{111},$ $\ket{+11},$ $\ket{++1},$ $\ket{+++}\}$ is shown in \cref{fig:simulation-4states}.

The periodicities of the input states can be determined from the location of the peaks of the QFTs, which are separated by $2^n/r$, where $r$ is the period and $n=3$ is the number of qubits. The separation of the peaks after the QFT corresponds to the number of peaks in the input state before the QFT, indicating the recurrence of the quantum amplitudes in a superposition of the eight computational basis states \cite{preskill2024}. 

The input states $\ket{111}, \ket{+11}$ and $\ket{++1}$ have periods $8$, $4$ and $2$, respectively. The state $\ket{+++}$ is an equal superposition state, so it has period $1$. These periods can be correctly estimated by identifying the peaks in the results of the QFT algorithm.
The histograms show that the simulation results generally agree with the theoretical predictions. The deviations are due to the approximations made during the derivation and optimization of the gate sequence. 
The metrics of performance present values close to $1$ for the four input states, so it can be concluded that the gate sequence \eqref{eq:Uopt} represents a good approximation of the QFT algorithm.

%%%%%%%%%%%%%%%%%%%%%%%%%%%%%%%%%%%%%%%%%%%%%%%%%%%%%%%%%%%%%%%%%%%%%%
\subsection{Noisy simulations}

Experimental imperfections in the setup introduce errors that affect the performance of the QFT. To simulate these effects, different noise models were used.

Coherent errors were introduced in the simulation with an ion-trap model of noisy phased rotations parameterized by an over-rotation error, $\varepsilon$, and a detuning error, $\delta$ \cite{noisygate}. In this model, \cref{eq:phasedrotation} becomes

\begin{equation}
    R_k^\text{noisy}(\theta, \phi) = e^{-i \frac{\theta}{2} (1+\varepsilon) (\sigma_x \cos \phi + \sigma_y \sin \phi) + \delta \frac{\theta}{2\Omega} \sigma_z}.
    \label{eq:phasedrotation-noisy}
\end{equation}

To introduce the stochastic nature of noise in the simulation of the QFT, the error parameters were considered random numbers within symmetric intervals, $\epsilon \in [-0.3, 0.3]$ and $\delta \in [-0.05, 0.05]$.

Through dephasing, caused in this setup by fluctuations of the magnetic field, the relative phase of the qubit superposition is randomized and the coherence decays exponentially over time \cite{khromova}. The dephasing channel can be expressed as

\begin{equation}
    \begin{split}
    \rho_{\text{dephased}}&(t) = e^{- \lambda t} \rho_0 + \\ &(1 - e^{- \lambda t}) \sum_i \ket{\psi_i} \bra{\psi_i} \rho_0 \ket{\psi_i} \bra{\psi_i},
    \end{split}
    \label{eq:dephasing}
\end{equation}
where $\rho_0$ is the three-qubit density matrix that represents the system before dephasing has occurred and $\ket{\psi_i} \in \{\ket{000},\ket{001},\ket{010},\ket{011},\ket{100},\ket{101},\ket{110},\ket{111}\}$ are the computational basis states. 
The decay constant, $\lambda$, was determined experimentally in \cite{versatile} with a value of $\lambda = 0.0625$ ms$^{-1}$. 
In the simulation, the dephasing channel was applied after each of the three periods of free evolution, with the duration of the process corresponding to the free evolution times, $T_1$, $T_3/2$ and $T_3/2$.

The final noise model that was included in the simulation was a depolarization process. Imperfect DD pulses and a limited readout efficiency cause a seemingly random behaviour of the qubits \cite{versatile}, which become depolarized, that is, the qubits fall into a completely mixed state. This effect can be expressed as

\begin{equation}
    \rho_{\text{depolarized}} = \frac{\zeta}{8} \mathbb{I} + (1- \zeta) \rho_{\text{dephased}},
    \label{eq:depolarization}
\end{equation}
where the phenomenological parameter $\zeta = 0.25$ describes the amount of white noise introduced in the final state.
In the simulation, the depolarizing channel was applied to the density operator of the final state of the QFT sequence implemented with the noisy rotations and dephasing models.

\begin{figure}[H]
    \centering
    \begin{minipage}{0.45\columnwidth}
        \centering
        \includegraphics[width=\linewidth]{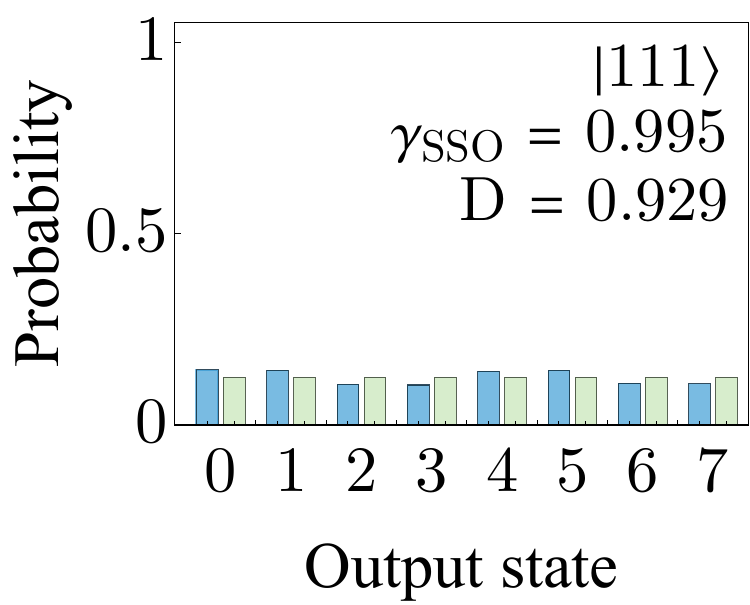}
    \end{minipage}
    \begin{minipage}{0.45\columnwidth}
        \centering
        \includegraphics[width=\linewidth]{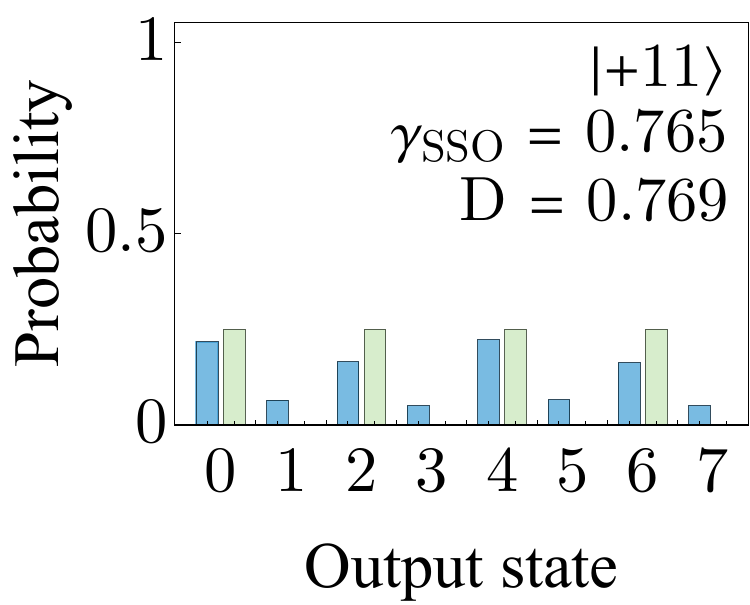}
    \end{minipage}
    \begin{minipage}{0.45\columnwidth}
        \centering
        \includegraphics[width=\linewidth]{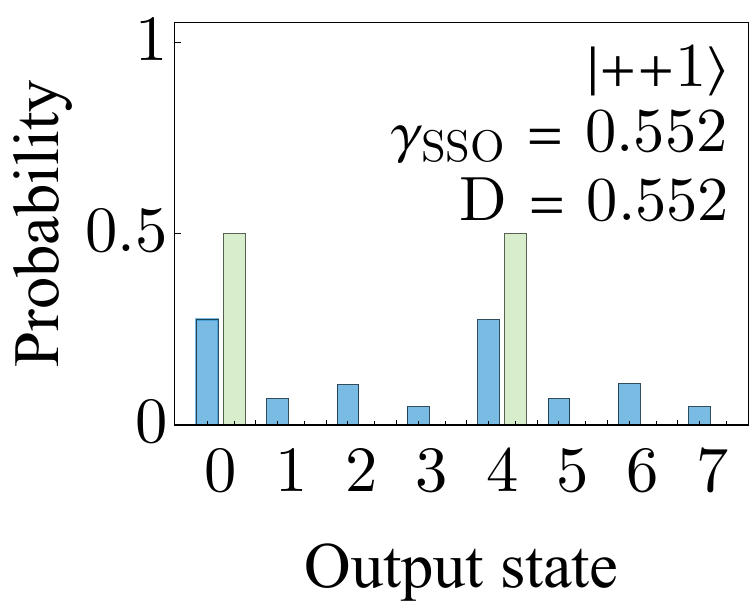}
    \end{minipage}
    \begin{minipage}{0.45\columnwidth}
        \centering
        \includegraphics[width=\linewidth]{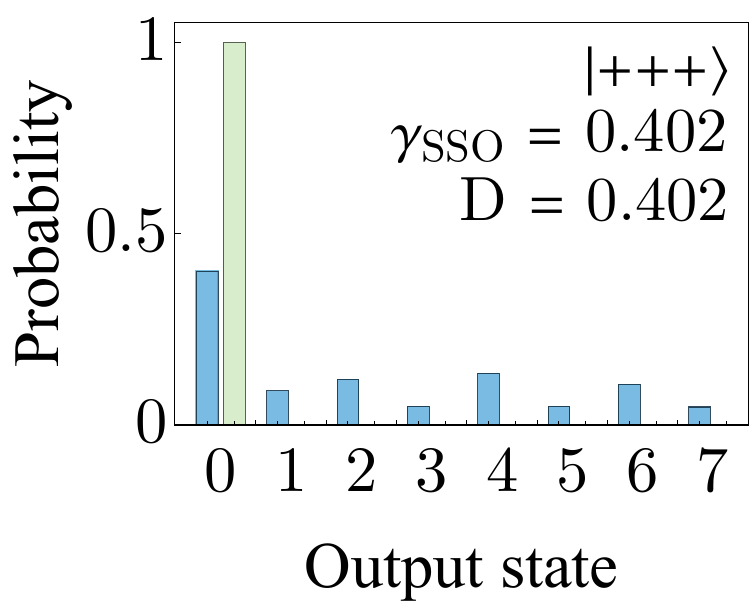}
    \end{minipage}
    \begin{minipage}{0.25\textwidth}
        \centering
        \includegraphics[width=\linewidth]{legend.pdf}
    \end{minipage}
    \caption{Results of the QFT simulation with the noisy gates, dephasing and depolarization models, for the four input states $\{\ket{111},$ $\ket{+11},$ $\ket{++1},$ $\ket{+++}\}$, showing the probability of each output state occurring in a measurement. The noise parameters of the model of noisy gates, the over-rotation error $\varepsilon$ and the detuning error $\delta$, were taken as random numbers. The dephasing decay constant is $\lambda=0.0625$ ms$^{-1}$ and the depolarization parameter is $\zeta=0.25$. The simulation results (blue) are compared with the theoretically expected results (green). The SSO $\gamma_{\text{SSO}}$ and distinguishability $D$ are metrics of the performance.}
    \label{fig:simulation-4states-depolarization}
\end{figure}

The results of the complete simulation of the QFT sequence, including the noise models for noisy gates, dephasing and depolarization, are shown in \cref{fig:simulation-4states-depolarization}. The simulation was repeated $1250$ times for each input state and the results refer to the mean of those simulations. The statistical error is too small to be shown.

\begin{figure}[H]
    \centering
    \begin{minipage}{\columnwidth}
        \centering
        \includegraphics[width=\linewidth]{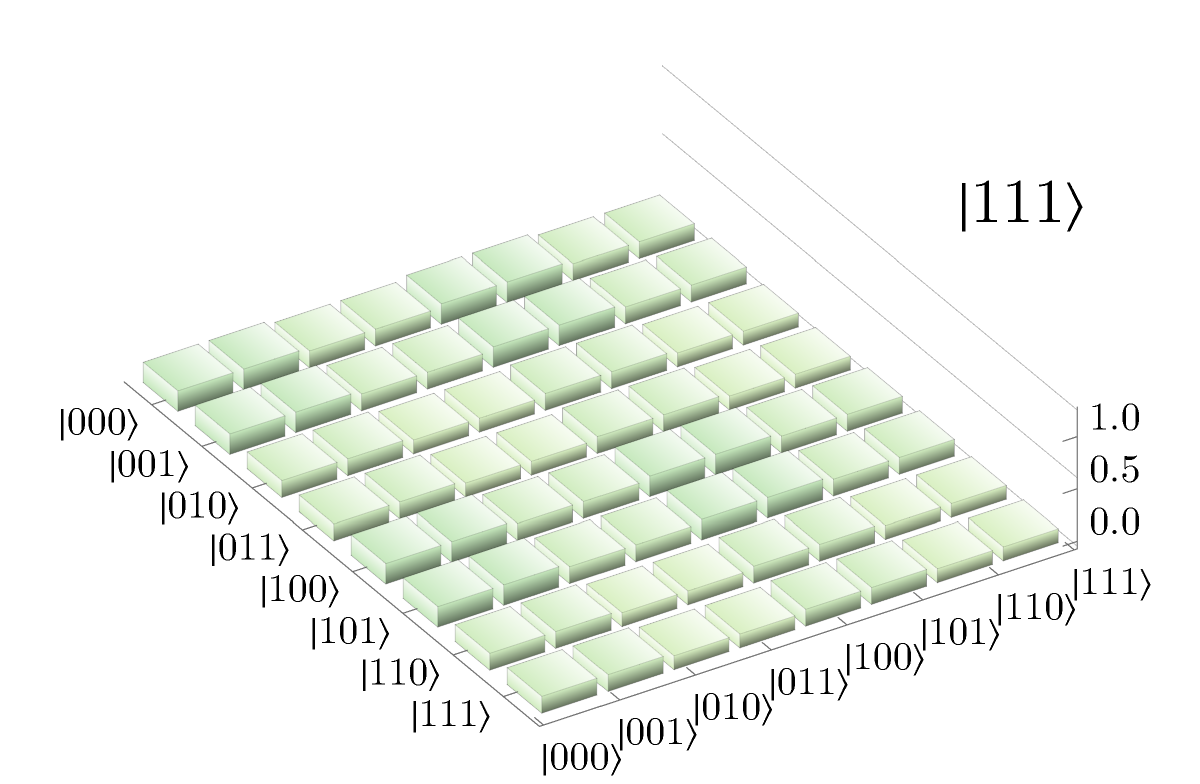}
    \end{minipage}
    \begin{minipage}{\columnwidth}
        \centering
        \includegraphics[width=\linewidth]{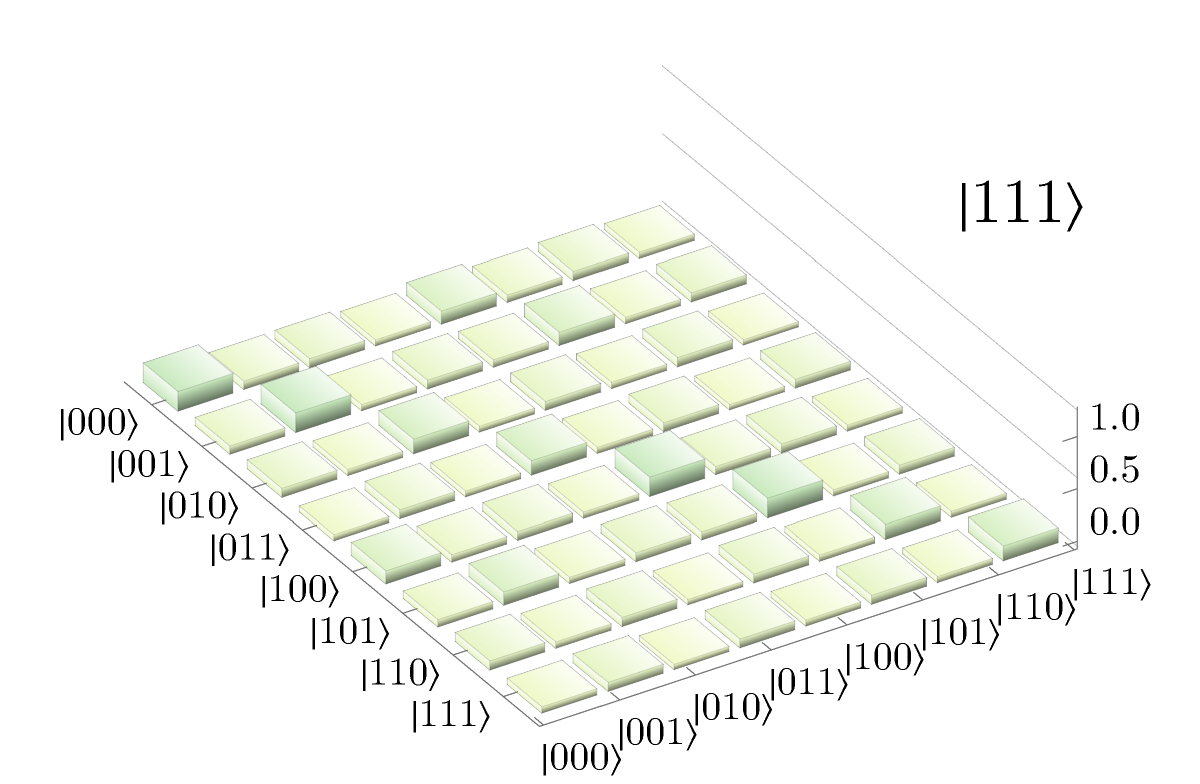}
    \end{minipage}
    \caption{Representation of the density operator corresponding to the final output state of the QFT, for the input state $\ket{111}$. Above are the results of the original simulation and below are the results of the noisy simulation.}
    \label{fig:simulation-densityoperator}
\end{figure}

In the noisy simulation, the performance of the algorithm is significantly decreased, compared to the original simulation. It can also be seen from the broad range of the metrics of performance that different input states have different susceptibilities to noise. 
However, the peaks of the QFT are still clearly identifiable for all input states. 
For instance, for the state $\ket{+++}$, $\gamma_{\text{SSO}}=0.953$ and $D=0.953$ without the noise models, and $\gamma_{\text{SSO}}=0.406$, $D=0.405$ with the noise models. 
For the state $\ket{111}$ the performance measures are now increased. This is deceitful because the output state of the simulation affected by noise is an almost completely mixed state, while in the original simulation without noise, a pure state in an equal superposition of the computational basis states is obtained. This distinction is clear in \cref{fig:simulation-densityoperator}, where the density operators of the output states of the QFT algorithm are represented. The density operators of both these states have a complete diagonal and since the chosen metrics of performance only look at the diagonal elements, they do not translate that differentiation.

\vspace{-1cm}

\vspace{-0.5cm}
The results of the noisy simulations demonstrate that the noise is an important component of quantum computation, with a large impact on the performance of the quantum algorithm.
\vspace{1cm}
%%%%%%%%%%%%%%%%%%%%%%%%%%%%%%%%%%%%%%%%%%%%%%%%%%%%%%%%%%%%%%%%%%%%%%
%     File: Paper_appendix.tex                               %
%     Tex Master: Paper.tex                               %
%                                                              
%%%%%%%%%%%%%%%%%%%%%%%%%%%%%%%%%%%%%%%%%%%%%%%%%%%%%%%%%%%%%%%%%%%%%%

%%%%%%%%%%%%%%%%%%%%%%%%%%%%%%%%%%%%%%%%%%%%%%%%%%%%%%%%%%%%%%%%%%%%%%

\section{Energetic Cost of a Microwave Pulse}
\label{appendix}

The qubits are manipulated by addressing the ions individually with microwave pulses. So, the energy required to perform the gates, as well as to apply the dynamical decoupling pulses, is given by the energy carried by the electromagnetic field of the microwave pulses \cite{sagar}.

The power carried by the electromagnetic field of the microwave pulse, $P$, is given by

\begin{equation}
    P(\omega) = \frac{1}{2} \frac{A_{\text{rad}}(\omega)}{A_{\text{dip}}}\frac{\hbar \Omega^2}{\cos^2\alpha},
    \label{eq:power}
\end{equation}
where $\omega$ is the microwave frequency, $\Omega$ is the Rabi frequency of the qubits and $\alpha$ is the angle between the magnetic moment of the ion and the incident magnetic field. To get a lower bound on the energy estimate, $\cos \alpha \approx 1$ was taken. $A_{\text{dip}} \approx 0.0034$ pm$^2$ can be thought of as an effective dipole cross-section for the ion and

\begin{equation}
    A_{\text{rad}}(\omega) = \frac{4 I_{11}}{(p_{11}')^2} \frac{\pi a^2}{\sqrt{1-(x(\omega))^2}}, \quad x(\omega) = \frac{c p_{11}'}{\omega a}
\end{equation}
is the effective area for the microwave pulse wavefront, dependent on the frequency of the microwave. Here, $I_{11} \approx 0.405$ is an integral of Bessel functions, $p_{11}' \approx 1.8412$ is the first root of the derivative of the first Bessel function $J_1'$ and $2a = 16.3$ mm is the diameter of the cylindrical cavity resonator that generates the microwave field.

The duration of each microwave pulse, $\tau$, is a function of the rotation angle $\theta$,

\begin{equation}
    \tau(\theta) = \frac{\theta}{\Omega}.
\end{equation}

Therefore, the energetic cost associated with each addressing microwave pulse depends on the resonance frequency of the ion that is being manipulated and on the rotation angle, 

\begin{equation}
    E(\omega,\theta) = P(\omega) \ \tau(\theta) = \frac{1}{2} \frac{A_{\text{rad}}(\omega)}{A_{dip}} \hbar \ \Omega \ \theta.
    \label{eq:energye}
\end{equation}

%%%%%%%%%%%%%%%%%%%%%%%%%%%%%%%%%%%%%%%%%%%%%%%%%%%%%%%%%%%%%%%%%%%%%%
\end{document}